\documentclass[12pt]{article}

\usepackage{amssymb}
\usepackage{amsmath}
\usepackage{graphicx}%
\usepackage{multirow}%
\usepackage{amsmath,amssymb,amsfonts}%
\usepackage{amsthm}%
\usepackage{mathrsfs}%
\usepackage[title]{appendix}%
\usepackage{xcolor}%
\usepackage{textcomp}%
\usepackage{manyfoot}%
\usepackage{booktabs}%
\usepackage{algorithm}%
\usepackage{algorithmicx}%
\usepackage{algpseudocode}%
\usepackage{listings}%
\usepackage{float}
\usepackage{mathrsfs}
\usepackage{rotating}
\usepackage{adjustbox}
\usepackage{booktabs} 
\usepackage{tabularx} 
\usepackage{array}    
\usepackage{url}    
\usepackage{natbib}

\title{Simulating Students with Large Language Models: A Review of Architecture, Mechanisms, and Role Modelling in Education with Generative AI}

\author{
  Luis Marquez-Carpintero\thanks{Corresponding author. Email: luis.marquez@ua.es}
  \and
  Alberto Lopez-Sellers\thanks{aj.lopez@ua.es}
  \and
  Miguel Cazorla\thanks{miguel.cazorla@ua.es}
}

\date{
  Institute for Computer Research \\
  University of Alicante \\
  San Vicente del Raspeig, P.O. Box 99, Alicante, Spain
}
\begin{document}
\maketitle 

\begin{abstract}
Simulated Students offer a valuable methodological framework for evaluating pedagogical approaches and modelling diverse learner profiles, tasks which are otherwise challenging to undertake systematically in real-world settings. Recent research has increasingly focused on developing such simulated agents to capture a range of learning styles, cognitive development pathways, and social behaviours. Among contemporary simulation techniques, the integration of large language models (LLMs) into educational research has emerged as a particularly versatile and scalable paradigm. LLMs afford a high degree of linguistic realism and behavioural adaptability, enabling agents to approximate cognitive processes and engage in contextually appropriate pedagogical dialogues. This paper presents a thematic review of empirical and methodological studies utilising LLMs to simulate student behaviour across educational environments. We synthesise current evidence on the capacity of LLM-based agents to emulate learner archetypes, respond to instructional inputs, and interact within multi-agent classroom scenarios. Furthermore, we examine the implications of such systems for curriculum development, instructional evaluation, and teacher training. While LLMs surpass rule-based systems in natural language generation and situational flexibility, ongoing concerns persist regarding algorithmic bias, evaluation reliability, and alignment with educational objectives. The review identifies existing technological and methodological gaps and proposes future research directions for integrating generative AI into adaptive learning systems and instructional design.
\end{abstract}




\section{Introduction}\label{sec1}

The role of the teacher is currently undergoing a profound yet quiet transformation, driven by the progressive integration of digital technologies in educational settings, a process that has been unfolding for years~\cite{davis2013restructuring,safidon2024educational}. While many institutions are still struggling to incorporate digital tools that have been available for over a decade, the recent emergence of Large Language Models (LLMs) introduces challenges of a fundamentally different nature. These models not only reshape classroom dynamics but also compel a complete reconsideration of what it means to teach and learn in AI-mediated educational contexts~\cite{Kozov2024Analyzing}.

Although the concept of simulating learners is not novel, early approaches, including rule-based frameworks and cognitive models, were hampered by the rigidity of hand-crafted interaction rules and their limited ability to capture the variability and spontaneity of human learning~\cite{vanlehn1994applications}. A systematic review published in 2023 addressed student simulation specifically but focused only on literature from 2010 to 2019, thus overlooking the recent wave of technological advancements enabled by generative models~\cite{kaser2024simulated}.

The capacity to simulate virtual students offers educators a low-risk means of experimenting with pedagogical strategies, curriculum design, and assessment methods. Such environments enable controlled evaluation before implementation in real classrooms, enhancing both scalability and pedagogical safety.

Nevertheless, several unresolved challenges remain. The degree to which LLMs can faithfully replicate human cognitive and affective processes is still under investigation. Concerns persist regarding algorithmic bias, limitations in open-access training datasets, and the risk of generating overly idealised or homogenised behaviours. Additionally, the diversity of real learners—encompassing differing personalities, backgrounds, and levels of ability—poses a significant challenge to the validity of these simulations, particularly when compounded by the biases embedded in training data.

This article addresses three core research questions:

\begin{itemize}
    \item What are the main cognitive architectures and implementation mechanisms (e.g., memory, reasoning) used to build Simulated Students based on LLMs?
    \item To what extent do LLM-based systems accurately simulate human behaviour in educational contexts?
    \item Can these simulations be effectively leveraged to improve teaching practices and instructional design?
    \item What are the main open research lines in this field? 
\end{itemize}

By reviewing recent empirical and methodological studies, this work provides an integrated synthesis of the current state of research on LLM-driven educational simulation. It outlines the technological advances, methodological constraints, and future directions for simulating students.

To our knowledge, this is the first dedicated literature review focused specifically on the simulation of student roles and mechanisms using LLMs, and we concentrate primarily on recent publications owing to the accelerated evolution of LLMs over the past few years~\cite{ho2024algorithmicprogresslanguagemodels}.

\subsection*{Outline}
The remainder of this survey is organised as follows. We first introduce foundational concepts before turning to architectural patterns, modelling mechanisms, evaluation practices, and applications. We conclude with open problems and implications for future work.
\textit{Section 2} introduces basic concepts and definitions of Simulated Students and related terminology.
\textit{Section 3} presents a unifying taxonomy and architectural patterns.
\textit{Section 4} reviews cognitive, affective, and trait modeling mechanisms.
\textit{Section 5} surveys task generation, feedback strategies, and evaluation protocols.
\textit{Section 6} summarizes applications, datasets, and benchmarks.
\textit{Section 7} synthesizes limitations and cross-cutting challenges.
\textit{Section 8} details \emph{Open Problems and Research Directions}.
\textit{Section 9} concludes with key takeaways and implications for future work.

\section{Theoretical Framework}

This section establishes the conceptual and technical framework for the application of LLMs in this field. The analysis is structured around two fundamental axes:
First, it traces the evolution of simulation systems, from pioneering rule-based approaches to current generative agents. Second, it delves into the theoretical foundations drawn from educational psychology and learning theories.
This dual technological and pedagogical pillar is essential for assessing the transformative potential of LLMs in educational simulation and for properly contextualising the findings of this review.

The Simulated Students or Simulated Learner paradigm in the pre-LLM era (1990–2019) focused on developing internal student representations for adaptive instruction within Intelligent Tutoring Systems (ITS)~\cite{harpstead19use}. Early approaches were primarily symbolic and rule-based, exemplified by systems like SCHOLAR, which employed semantic nets to track known concepts~\cite{nwana1990intelligent}. These systems, which gained prominence in the 1980s and 1990s, were designed to provide individualized and immediate feedback, and to adapt instructional content based on a dynamically maintained model of student performance and progress \citep{nwana1990intelligent}.

The most ambitious goal of this period was the development of cognitive models, with Anderson’s LISP Tutor being the landmark example based on the ACT-R architecture~\cite{ritter2019act, vantsent, anderson1990cognitive}. These systems utilised Model-Tracing to diagnose the student's problem-solving trajectory against an Ideal Student Model. While offering high fidelity and explainability, these models suffered from critical limitations: the high cost of intensive knowledge engineering and the inherent in-scalability required to anticipate and catalogue every potential error (mal-rule or bug).

To overcome the scaling limitations of symbolic systems, the field shifted toward probabilistic and data-driven methods. Bayesian Knowledge Tracing (BKT), introduced by Corbett and Anderson (1994), became the canonical approach, using a Hidden Markov Model (HMM) to statistically estimate skill mastery over time~\cite{shchepakin2023parametric}. Additionally, BKT defines four core parameters—Prior, Learning (Transition), Guess, and Slip—to infer the latent knowledge state, and it remains a relevant foundation in the field of student knowledge Modelling~\cite{shen2021learning}.

Subsequently, starting around 2000 and growing through the 2010s, analytics-based approaches emerged, leveraging real student interaction data. This period saw the rise of Educational Data Mining (EDM)~\cite{dutt2017systematic, romero2007educational} and Learning Analytics (LA)~\cite{chatti2012reference}, which applied data mining and supervised Machine Learning to educational processes. These methods focused on the prediction of academic outcomes, the early detection of at-risk students~\cite{huang2020predicting}, and identifying behavioural patterns to predict performance, learning difficulties, and potential dropouts, particularly in response to the rise of e-learning~\cite{dutt2017systematic, romero2007educational}.

Simultaneously, beyond knowledge, researchers began to model cognitive and affective aspects of learning, addressing emotions, motivation, and other affective aspects using fuzzy logic, Bayesian networks, or clustering techniques~\cite{chrysafiadi2013student, bakhshinategh2018educational}.

Despite these advances, a systematic review covering the 2010–2019 literature by~\cite{kaser2024simulated} critically summarized the era’s shortcomings, finding that Simulated Students models tended to represent only “narrow aspects of student learning”. Most strikingly, the review found that almost half of the simulated learner studies failed to provide any formal validation of their simulations. This crisis of fidelity and validation marked the end of the pre-LLM period, highlighting the compromises made between technical viability and psychological realism.

\subsection{Foundations in Educational Psychology and Learning Theories}

The simulation of students leveraging LLMs, a remarkable achievement in AI, explores the feasibility of simulating student learning behaviours. Unlike conventional machine learning-based prediction, LLMs are utilised to instantiate virtual students with specific demographics and uncover intricate correlations among learning experiences, course materials, understanding levels, and engagement. The objective is not merely to predict learning outcomes but to replicate the learning behaviours and patterns of real students. Collectively, these findings deepen the understanding of LLMs and demonstrate their viability for student simulation, thereby empowering more adaptable curricula design to enhance inclusivity and educational effectiveness~\cite{xu2023leveraging}.

From the constructivist perspective, learning is understood as an active process in which the student constructs knowledge through interaction with the environment~\cite{prawat1994philosophical}. 
LLM-based models can incorporate these principles by analyzing and categorising their own learning capabilities, which can be analogous to varying levels of prior knowledge. In particular, the concept of the Zone of Proximal Development (ZPD) is useful for designing adaptive interventions within in-context learning (ICL) and fine-tuning scenarios for LLMs, where demonstrations or training examples serve as scaffolding or gradual assistance based on the LLM's own 'cognitive state'~\cite{cui-sachan-2025-investigating}.

On the other hand, cognitive learning theories, such as information processing theory, offer tools for Modelling cognitive load, attention, working memory, and metacognitive processes. These elements have begun to be integrated into the architectures of simulated educational students, such as Agent4Edu~\cite{gao2025agent4edu} and Classroom Simulacra~\cite{xu2025classroom}, where virtual learners exhibit dynamic cognitive states and respond differently depending on their level of conceptual mastery.

\section{Methodology}~\label{methodology}

This study employs a thematic review approach to explore and synthesize recent advances in the use of LLMs for student simulation in educational contexts, offering a complementary perspective to existing thematic reviews. To ensure a rigorous and transparent process, this section describes the steps taken in detail. It presents the systematic search strategy and selection criteria~\ref{search_strategy}, followed by a discussion of the limitations of this approach~\ref{limitation_search_strategy}.

\subsection{Search Strategy}\label{search_strategy}
This study employs a systematic search strategy to explore and synthesise recent advances in the use of LLMs for student simulation in educational contexts. The findings were synthesised using a thematic approach.

A systematic search was conducted in the Scopus, Web of Science, IEEE Xplore, ACM Digital Library, and Google Scholar databases. Results from Google Scholar were incorporated to identify recent publications that might not yet have been indexed in the main databases or retrieved in the initial search.

Some of the search terms used include:
\begin{itemize}
\item Simulated Students
\item educational simulation
\item digital twin
\item generative agents in classrooms
\item student Modelling with LLMs
\end{itemize}

The selection of studies focused primarily on peer-reviewed articles relevant to the field, prioritising those published between 2021 and 2025, a pivotal period in the expansion of LLMs. Given the field’s rapid evolution and practical relevance, significant preprints were also included.

The identified articles were evaluated based on thematic relevance. Studies were included if they met at least one of the following criteria:
\begin{itemize}
    \item Methodological proposals or system architectures aimed at simulating students using LLMs.
    \item Experimental or conceptual applications of LLMs in simulated educational environments.
    \item Studies addressing cognitive, affective, or personality dimensions in generative educational agents.
\end{itemize}

Given the narrative nature of the review, no restrictions were applied regarding the type of study (theoretical, technical, or experimental); instead, the objective was to represent a diversity of approaches and perspectives.

\begin{figure}[H]
    \centering
\includegraphics[width=0.55\linewidth]{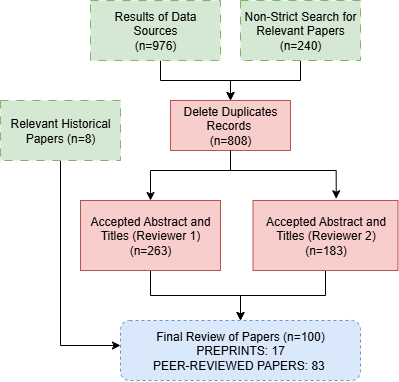}
    \caption{PRISMA flow diagram illustrating the process of study identification, screening, eligibility, and inclusion for this review.}
    \label{fig:prisma}
\end{figure}

Figure~\ref{fig:prisma} presents the PRISMA flow diagram, which details the study selection process. This process was undertaken by two independent reviewers to ensure the reliability of the selection and consistency in the application of the criteria. Following the removal of 408 duplicates, the two reviewers filtered 808 titles. Discrepancies arising at this stage were resolved through discussion and consensus. Finally, articles of historical relevance were manually incorporated, even if they did not strictly meet the pre-established inclusion criteria, with the objective of adequately contextualising the research area.

\begin{figure}[h!]
    \centering
\includegraphics[width=0.8\linewidth]{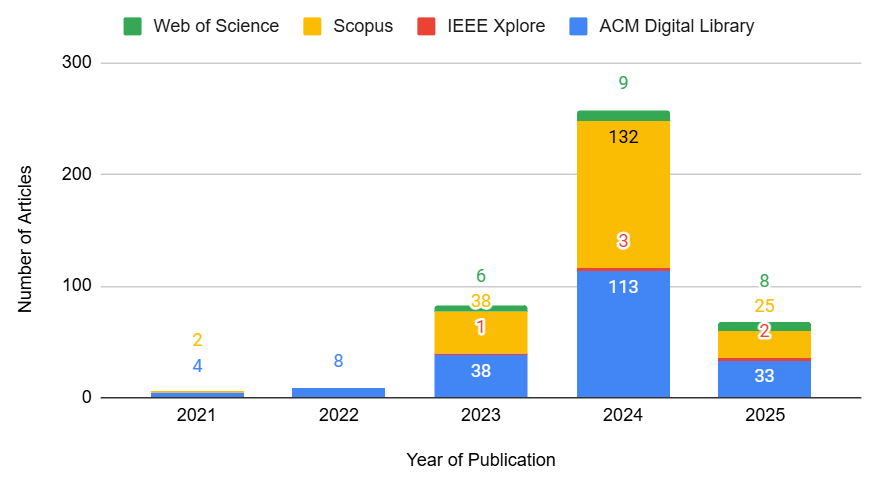}
    \caption{Evolution of the number of publications on Simulated Students.}
    \label{fig:evolution}
\end{figure}

As shown in Figure~\ref{fig:evolution}, the number of articles on LLMs in education was collected across various journals and search platforms, including Google Scholar, Scopus, and others, highlighting an emerging field that uses models such as ChatGPT or LLaMA to simulate students in educational contexts, enabling not only the modelling of individual behaviour but also the dynamics of learning within entire academic ecosystems. Google Scholar was omitted from the figure because we manually extracted a fixed number of papers from it to ensure the relevance of the sources.

The selected articles were read in depth, and findings were thematically organised. Rather than quantifying results, a qualitative synthesis was chosen to identify patterns, conceptual tensions, and emerging lines of research. Additionally, to enhance the accuracy and rigour of this article, a double-verification process was conducted using the Gemini and ChatGPT LLMs.

\subsection{Limitations of the Search Strategy}\label{limitation_search_strategy}

The data were synthesized thematically, identifying common trends in student simulation with LLMs, limitations of existing approaches, and potential areas for improvement.

Although a search strategy was employed to identify relevant literature on the simulation of students, teachers, and classrooms using LLMs, it is important to acknowledge several inherent limitations of this approach. The limitations of this study are:

\begin{itemize}
    \item The possibility of overrepresentation of studies with positive results.
    
    \item The review was limited to studies published exclusively in English to ensure the academic quality and impact of the included publications.
    
    \item Finally, the review did not include so-called grey literature, such as master's or doctoral theses, unpublished institutional documents, developers' technical blogs, or content from communities of practice.
\end{itemize}

Taken together, these limitations do not invalidate the findings of the review but rather define the scope and potential omissions of the analysed corpus. Acknowledging them allows for a critical reading of the findings and reinforces the commitment to methodological transparency in the development of this thematic review.

\section{Cognitive Modelling}

Realistically simulating students with LLMs requires going beyond generating textual responses. Incorporating cognitive and affective factors—such as knowledge level, confusion, interest, or motivation—enables the creation of more believable agents that are useful for evaluating pedagogical strategies or training teachers in diverse scenarios.

This section analyses current approaches for representing these states, the theoretical frameworks that support them, and the technical mechanisms used to implement them, such as contextual memory or iterative reflection. It also discusses their limitations and the potential to enhance personalization and adaptability in simulated educational environments.

\subsection{Architecture and Implementation Mechanisms}

Recent works in the literature explore approaches that optimise architectural design, including multi-agent systems, modular components, and the integration of diverse generative agents to produce more robust educational simulations. While these proposals vary in complexity, they share a common goal: to realistically represent learning behaviours using LLM-powered generative agents.

\begin{figure}[H]
    \centering
    \includegraphics[width=1\linewidth]{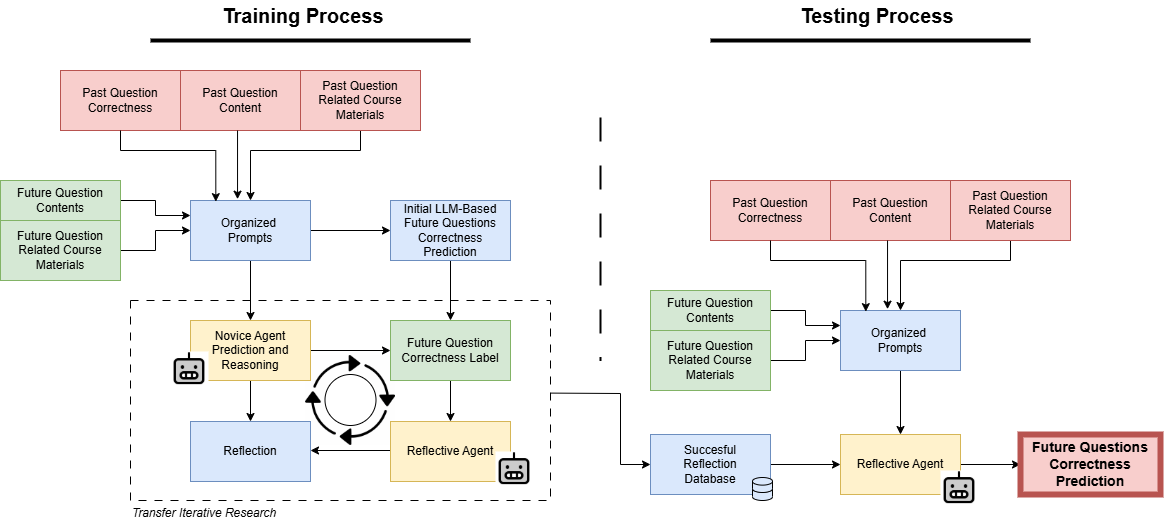}
    \caption{Reflective mechanism of the TIR module.}
    \label{fig:tir}
\end{figure}

One notable contribution is Classroom Simulacra \citep{xu2025classroom}, which focuses on simulating learning behaviour with LLMs. Figure \ref{fig:tir} presents its Transferable Iterative Reflection (TIR) module, designed to enhance LLM performance through prompting and fine-tuning. In brief, during training a reflective agent issues an initial prediction, compares it with ground truth, writes a short reflection, and hands that to a novice agent that re-predicts; the loop repeats until accuracy plateaus, and the most helpful reflections are stored in a repository. During testing, the model retrieves a small set of those stored reflections, combines them with the student’s history, and predicts future answers without ever seeing the true labels. In this framework, course knowledge is delivered via lecture slides, which serve as external stimuli, while prior knowledge acts as internal stimuli. The model evaluates its ability to capture variation in post-test accuracy at multiple levels (individual, lecture, question, and skill), showing that knowledge is primarily transferred through exposure to course materials, and the TIR reflections refine this representation to improve behavioural simulation.

\begin{figure}[h!]
    \centering
    \includegraphics[width=0.85\linewidth]{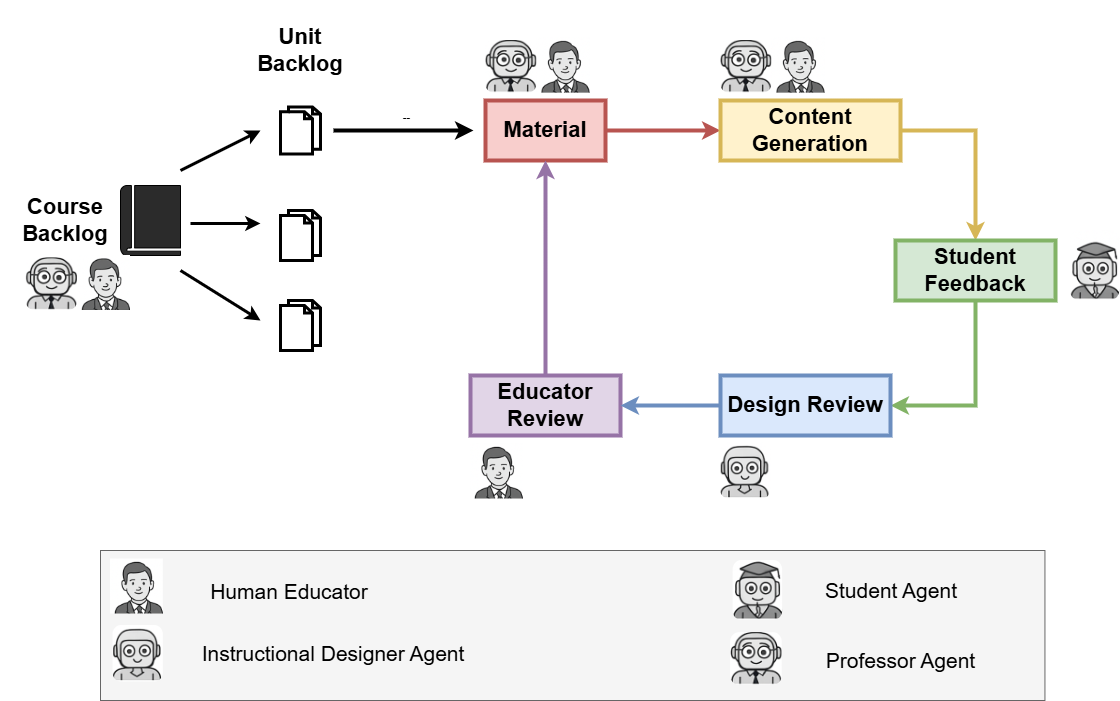}
    \caption{Cyclic communication model in the AICademic multi-agent system.}
    \label{fig:wip_cycle}
\end{figure}

Further advancing multi-agent interaction, WIP: Active Learning Through Prompt Engineering and Agentic AI Simulation, A Pilot Project in Computer Networks Education \citep{ma2024wip} introduces a system with three primary agent roles: teacher, student, and instructional designer. As shown in Figure \ref{fig:wip_cycle}, the AlCademic platform sets up a collaborative cycle where the human educator first decomposes the course into a backlog of learning outcomes, selects a subset to form the unit backlog for a single lecture, and then allows the professor, student, and instructional designer agents to iterate through content generation, student feedback collection, design review, educator validation, and material updating until the unit objectives are met.

Another notable approach is Agent4Edu~\cite{gao2025agent4edu}, which introduces a personalized learning simulator and incorporates reflective mechanisms within its memory module. This system employs LLM-based generative agents to simulate student response data and detailed problem-solving behaviour. Agents are equipped with predefined student profiles, memory modules, and action modules designed specifically for personalized learning scenarios. These agents are capable of selecting, understanding, analyzing, and responding to exercises in a human-like manner, which allows for more accurate prediction of student responses.

Finally, the SimClass framework~\cite{zhang2024simulating} offers another multi-agent architecture for simulating real-time interactions between teachers and students. The framework identifies key classroom roles and designs a class control mechanism. Courses are deployed using prepared scripts and lecture slides as the basis for agent interactions. SimClass features several agent types: a Teacher Agent responsible for delivering content and answering questions, an Assistant Agent that supplements instruction, participates in discussions, maintains discipline, and enhances learning efficiency, and Classmate Agents with defined personality traits to simulate peer dynamics. Users may also customize additional agents. The evaluation of this system employs the Flanders Interaction Analysis System to examine the teaching style of the agents.

\subsection{Memory Management}

In the field of educational agents powered by LLMs, memory management, information access, and contextual recall are fundamental components for realistically simulating human learning processes. A variety of recent studies explicitly address these mechanisms, highlighting technical advances in how these virtual entities store, retrieve, and reflect upon information in educational contexts.

Several works tackle the concept of memory and information retrieval in the context of Simulated Students, employing different approaches to enable more realistic and adaptive interactions. In Agent4Edu~\cite{gao2025agent4edu}, a memory management system is introduced for simulated agents. This system, includes three main operations. The first operation, memory retrieval, enables an agent to extract pertinent information from both short- and long-term stores, recognize reinforced facts and produce succinct summaries. The second operation, memory writing, initially records raw observations as factual entries and, upon reinforcement, promotes them into short- or long-term memory. The third operation, memory reflection, occurs exclusively in long-term memory and takes two forms: summary reflection, which integrates accumulated memories into high-level insights, and corrective reflection, which is triggered when the agent’s actions deviate from those of an actual student. This architecture emphasizes that traditional simulators are often limited to short-term memory, whereas combining factual response records (specific) with learning memory summaries (general) enables more accurate Modelling of student practice.

In \textit{Classroom Simulacra: Building Contextual Student Generative Agents in Online Education for Learning behavioural Simulation}~\cite{xu2025classroom}, a process known as Time-aware Iterative Reflection (TIR) is described. This mechanism allows generative agents to learn from incorrect predictions and build a successful reflection database for each lesson. This database can be considered a form of long-term memory that directly influences future decision-making by the agent. The reflection process is performed offline for each lesson, using pre-prepared materials, suggesting a deliberate memory consolidation strategy.

\begin{figure}
    \centering
    \includegraphics[width=1\linewidth]{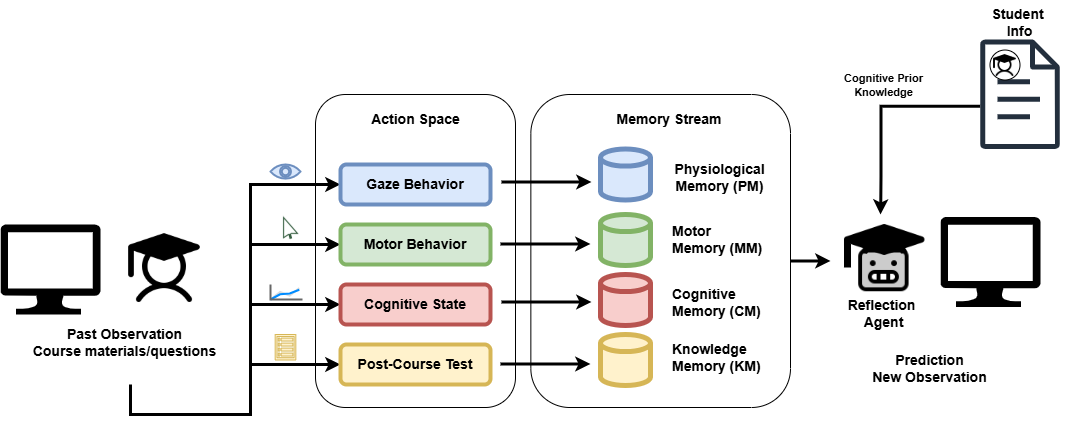}
    \caption{EduAgent framework pipeline.}
    \label{fig:eduagentimg}
\end{figure}

The EduAgent: Generative Student Agents in Learning framework described in a preprint by~\cite{xu2024eduagent} is illustrated in Figure~\ref{fig:eduagentimg}. The figure is a block diagram that shows how each new slide triggers four parallel outputs, gaze, motor action, cognitive state and test response, each one stored in its own memory before being retrieved for reflection and the next step. The framework underscores the importance of memory in simulating realistic learning behaviours. Agents instantiated with personality profiles simulate the learning process slide by slide. Before executing any action, such as gaze shifts, motor control, cognitive state transitions or responses, the agents reflect on the relationship between their character, past actions and previous course materials stored in memory, relying on integrated cognitive priorities. These demonstrations provide a temporal sequence of past behaviours and course materials, allowing the agent to reason about the interactions between different behavioural types. Notably, only behaviours from the most recent slide are used as demonstrations, and ablation studies showed that removing past cognitive states from memory significantly reduced performance, highlighting the critical role of contextual memory.

Following this direction, the preprint MathVC~\cite{yue2024mathvc} reinforces the necessity of both short- and long-term memory. Here, short-term memory is implemented via a dynamic dialogue history, allowing agents to maintain coherence and conversational continuity. Long-term memory, on the other hand, is represented through a symbolic structure called a character schema, which stores the student's evolving understanding of the mathematical problem under discussion. This schema is incrementally updated during the interaction, Modelling initial errors, revisions, and corrections analogously to real human cognition. A meta-planner organizes dialogue into structured collaborative stages, supporting both individual knowledge development and group coordination, thereby reinforcing the educational value of structured memory in generative agents.

In \textit{Fuzzy Memory Networks and Contextual Schemas: Enhancing ChatGPT Responses in a Personalized Educational System}~\cite{troussas2025fuzzy}, the focus is on simulating human-like teaching behaviour through an AI-driven tutor rather than Modelling student behaviour directly. The system proposes a forgetting-based approach to enhance ChatGPT’s role as a simulated tutor, enriching its responses with fuzzy logic-selected memories, thematically relevant clusters, and specific learning history data. This architecture reflects how human memory operates, using fuzzy membership functions and adaptive thematic clustering to produce context-aware, personalized responses. Although the student is not directly modelled, the system simulates their presence through stored past interactions, such as questions, comments, and feedback, which serve as memory traces processed based on recency, relevance, and interaction type.
The system is composed of three core modules: the Fuzzy Memory Module, which simulates memory retention through fuzzy logic; the Schema Manager, which organizes interactions into thematic clusters; and the Response Generator, which crafts personalized responses based on prioritized memory traces.

In contrast to this personalized feedback-oriented approach, a more general and widely adopted method in other roles is presented in \textit{Generative Agents: Interactive Simulacra of Human behaviour}~\cite{park2023generative}. This work introduces behaviour simulation based on continuous memory generation and updating. For example, a character such as “Mei”, initially defined as a professor interested in research and teaching at 9 a.m., behaves in a manner consistent with this background while continuously enriching her memory through interactions with the environment, including activities like teaching and conversing with colleagues. Future actions are conditioned on this memory history. No pedagogical feedback is provided; instead, the agent behaves as a person with specific occupations and traits. A similar process applies to Simulated Students. For instance, “Eddy”, characterized as a music student, forms plans such as “attend class” or “practice compositions” and reacts to contextual events like study invitations. This active memory-based simulation enables Modelling both spontaneous behaviour and decision-making shaped by prior experiences.

\subsection{Knowledge Levels}

The representation of cognitive levels in Simulated Students adopts various approaches grounded in theoretical frameworks from educational psychology. A common method involves the use of knowledge components (KCs) to model a student’s mastery over specific concepts—such as demonstrated understanding, confusion, or insufficient evidence—which allows for the simulation of learners with targeted gaps or misconceptions~\cite{jin2024teachtune}. Building on these fine-grained KC profiles, Lu and Wang introduce Generative Students, a framework that samples mastered, confused and unknown KCs to craft realistic learner personas~\cite{lu2024generative}. Additionally, students have been simulated with varying levels of general ability (from low to high), which has proven useful for assessing the effectiveness of evaluation items and informing tutoring strategies~\cite{benedetto2024using}.

The ability to represent and structure knowledge is essential in educational environments and intelligent systems. Accordingly, several methodologies have emerged to capture, organize, and utilize information efficiently, supporting accurate simulation or transfer of knowledge. Within this review, three clearly differentiated strategies are identified for Modelling student knowledge, which are summarized in Table~\ref{tab:knowledge_Modelling_comparison}.
\renewcommand{\arraystretch}{1.5} 
\begin{table}[ht!]
\centering
\small
\resizebox{\textwidth}{!}{%
    \begin{tabular}{|p{2.5cm}|p{5cm}|p{5cm}|p{5cm}|}
\hline
\textbf{Strategy} & \textbf{General Description} & \textbf{Advantages} & \textbf{Limitations} \\
\hline
\textbf{Direct \newline Simulation \newline via Prompts}  & Knowledge levels, common mistakes, or learning styles are directly defined within the LLM prompt.   & - Easy to implement \newline - Highly customizable \newline - Useful for evaluating questions and strategies  & - Relies heavily on prompt quality \newline - May lead to overly idealized behaviours \\
\hline

\textbf{Knowledge \newline Tracing}  & Tracks the mastery of knowledge components over time using models such as DKT, DKVMN, AKT, or LLM-KT.  & - Strong predictive performance \newline - Enables real-time adaptation \newline - Integrates with deep learning and LLMs  & - Technically complex \newline - Requires large datasets \newline - Interpretation can be challenging \\
\hline

\textbf{Knowledge Graphs and Heuristics}  & Structured representations (KGs) and explicit rules define the student's knowledge state.  & - High interpretability \newline - Supports symbolic personalization \newline - Incorporates feedback and context   & - Requires extensive manual curation \newline - Struggles with dynamic learning processes \\
\hline
\end{tabular}
}
\caption{Comparison of knowledge modelling strategies for Simulated Students.}
\label{tab:knowledge_Modelling_comparison}
\end{table}

\subsubsection{Direct Prompt-Based Simulation with LLMs}

The most widely employed approach in the literature involves directly encoding knowledge levels within the prompt. This technique allows the instantiation of virtual student agents with differentiated profiles based on prior knowledge, mastered or confused concepts, and learning styles.

EduAgent~\cite{xu2024eduagent} introduces a framework for the realistic simulation of learning behaviours in online education. Each virtual student agent is instantiated with a persona profile and simulates the learning process through lecture slides. At each step of the simulation (i.e., each slide), the agent receives the transcript via a prompt and generates a trajectory of simulated learning behaviours. Before acting, agents reflect on the correlation between their persona, past actions, and previous course materials stored in memory, guided by integrated cognitive priorities. These demonstrations provide a time-series format of prior gaze/motor behaviours, cognitive states, and course materials, enabling the agent to reason about how different behaviours influence each other and apply reflective insights to new instructional content or test questions. This again emphasizes a memory- and reflection-based approach as the primary mechanism for knowledge transfer and Modelling.

A recent study applies Item Response Theory (IRT) to evaluate whether LLMs can reliably simulate human responses and calibrate educational items~\cite{liu2025leveraging}. Six models were tested: GPT-3.5, GPT-4, LLaMA 2, LLaMA 3, Gemini-Pro, and Cohere Command R Plus. Each was prompted with a single input containing 20 university-level algebra questions from the OpenStax textbook, formatted as “Q1: ..., Q2: ..., ..., Q20: ...”. The generated responses were manually scored and converted into binary vectors (correct/incorrect). These were analysed using IRT to compare the estimated ability of the models against real students. While some LLMs, such as GPT-3.5, demonstrated high performance, they exhibited narrower zones of ZPD, limiting their capacity to reflect human variability. However, combining human and synthetic responses at a 1:1 ratio resulted in more accurate item calibration, suggesting that LLMs can serve as an effective complement in educational assessment.

Some approaches focus explicitly on Modelling learning difficulties by encoding them directly into prompts. \textit{Using LLMs to Simulate Students' Responses to Exam Questions}~\cite{benedetto2024using} centers on using LLMs to simulate student responses to exam questions, equipping Simulated Students with varying proficiency levels. Knowledge is conveyed to the LLM via the exam question and contextual information in the prompt. The study concludes that an LLM’s ability to simulate students with different levels of understanding depends on well-crafted prompts that guide the model to use its pre-trained knowledge and supplied context to replicate domain comprehension. While not proposed within this particular study, it is recognized that for enhanced simulation fidelity, integrating retrieval-augmented generation (RAG)~\cite{lewis2020retrieval} with topic-specific documents could refine the definition of a Simulated Student's proficiency level.

TeachTune \citep{jin2024teachtune} employs a Personalized Reflect-Respond LLM-prompting pipeline that simulates student behaviour through three consecutive prompts: Interpret converts the instructor-specified knowledge components and trait ratings into a concise learner profile, Reflect updates this knowledge state after each turn, and Respond generates the student's next utterance conditioned on the updated state, trait profile, and dialogue history.

\subsubsection{Using Knowledge Tracing}
For a simulated agent to be pedagogically useful, it is not enough for it to generate fluent text; it must possess an internal cognitive model that simulates the acquisition, forgetting, and application of knowledge. This mechanism is what allows the simulator to learn from instruction, make realistic mistakes, and display a coherent learning trajectory.

The field of Knowledge Tracing (KT) is the established computational approach, predating LLMs, designed specifically to model this dynamic: it tracks a student's knowledge state over time. Therefore, the techniques and architectures developed for KT are fundamental, as they provide the foundation for constructing the cognitive engine of student simulators.

In its classic form, KT is a well-established approach that uses KCs (Knowledge Components) to track and predict student performance over time. It works by dynamically estimating the level of mastery of each individual KC, enabling systems to anticipate student success in future tasks and adapt instruction accordingly. Amongst recent advances, the KCGen-KT model~\cite{duan2025automated}, outlined in a recent preprint, extends beyond binary prediction of correct or incorrect answers by incorporating code generation conditioned on the student’s mastery profile across relevant KCs.

Recent approaches in KT have also introduced contextual enhancements. Some models now consider additional signals such as student response time, enabling a more fine-grained analysis of performance and learning behaviour~\cite{Zhou2025AAKT}.

To support and standardize experimentation in this space, the PyKT framework~\cite{liu2022pykt} offers a unified platform that integrates ten of the most commonly used deep learning KT (DLKT) models and seven public datasets. A comparative performance analysis of these models, as reported in the original study, is presented in Table~\ref{tab:rendimiento}.

\vspace{1em}
\renewcommand{\arraystretch}{1.3}
\begin{table}[ht]
\centering
\small
\resizebox{\textwidth}{!}{%
\begin{tabular}{lccccccccccccccc}
\toprule
\textbf{Model} & \multicolumn{2}{c}{Statics2011} & \multicolumn{2}{c}{AS2009} & \multicolumn{2}{c}{AS2015} & \multicolumn{2}{c}{AL2005} & \multicolumn{2}{c}{BD2006} & \multicolumn{2}{c}{NIPS34} & \multicolumn{2}{c}{POJ} \\
& L & S & L & S & L & S & L & S & L & S & L & S & L & S \\
\midrule
DKT    & 0.8219 & 0.8314 & 0.7351 & 0.7650 & 0.7106 & 0.7281 & 0.8160 & 0.7623 & 0.8010 & 0.8563 & 0.7740 & 0.7430 & 0.5979 & 0.6629 \\
DKT+   & 0.8276 & 0.8364 & 0.7357 & 0.7657 & \textbf{0.7113} & \textbf{0.7296} & 0.8168 & 0.7600 & 0.8015 & 0.8593 & 0.7748 & 0.7436 & 0.6045 & 0.6782 \\
DKT-F  & 0.7859 & 0.7465 & - & - & - & - & 0.8158 & 0.7597 & 0.7980 & 0.8467 & 0.7784 & 0.7480 & 0.5915 & 0.6606 \\
KQN    & 0.8230 & 0.8280 & 0.7259 & 0.7604 & 0.7064 & 0.7266 & 0.8038 & 0.7466 & 0.7931 & 0.8515 & 0.7738 & 0.7414 & 0.5944 & 0.6774 \\
DKVMN  & 0.8086 & 0.8294 & 0.7271 & 0.7588 & 0.7039 & 0.7240 & 0.8067 & 0.7429 & 0.7978 & 0.8540 & 0.7725 & 0.7414 & 0.5924 & 0.6732 \\
ATKT   & 0.8046 & 0.8295 & 0.7249 & 0.7605 & 0.7029 & 0.7262 & 0.8004 & 0.7564 & 0.7884 & 0.8464 & 0.7711 & 0.7438 & 0.5960 & 0.6687 \\
GKT    & 0.8044 & 0.8004 & 0.7224 & 0.7535 & 0.7111 & 0.7266 & 0.8122 & 0.7528 & 0.8042 & 0.8535 & 0.7741 & 0.7431 & 0.5977 & 0.6577 \\
SAKT   & 0.7958 & 0.8179 & 0.6989 & 0.7403 & 0.6857 & 0.7134 & 0.7891 & 0.7347 & 0.7734 & 0.8239 & 0.7570 & 0.7253 & 0.6001 & 0.6544 \\
SAINT  & 0.7592 & 0.7845 & 0.6687 & 0.7112 & 0.6617 & 0.7060 & 0.7788 & 0.7097 & 0.7776 & 0.8189 & 0.7912 & 0.7687 & 0.5294 & 0.6702 \\
AKT    & \textbf{0.8305} & \textbf{0.8466} & \textbf{0.7781} & \textbf{0.7878} & \textbf{0.7113} & 0.7292 & \textbf{0.8317} & \textbf{0.7771} & \textbf{0.8204} & \textbf{0.8643} & \textbf{0.8074} & \textbf{0.7829} & \textbf{0.6137} & \textbf{0.6949} \\
\bottomrule
\end{tabular}
}
\caption{Performance of KT models on different datasets (L: Long Interactions, S: Small Interactions), as presented in \cite{liu2022pykt}.}
\label{tab:rendimiento}
\end{table}

Beyond performance benchmarking,~\cite{zanellati2024hybrid} highlights several key models that have shaped the evolution of KT:

\begin{itemize}
    \item \textbf{DKT (Deep Knowledge Tracing)}: Introduced RNNs for KT, but suffers from temporal instability and limited interpretability.
    \item \textbf{DKVMN (Dynamic Key-Value Memory Networks)}: Improves tracking of individual concepts, although it neglects student behavioural factors.
    \item \textbf{Graph-based models}: Explicitly model relationships between exercises and concepts using Graph Neural Networks (GNNs).
    \item \textbf{Hybrid models}: Integrate prior knowledge into embeddings or loss functions. 
\end{itemize}

This progression culminates in the integration of LLMs into hybrid KT architectures. This is where the bridge to simulation becomes explicit: while models like LLM-KT, which is detailed in a preprint~\cite{wang2025llm}, are evaluated on predictive tasks (outperforming classical models as shown in Table~\ref{tab:resultllmkt}), their architecture (an LLM conditioned by a KT-derived state) is a direct precursor to a generative simulator.
By shifting the task from predicting a label (correct/incorrect) to generating a response (an explanation, a solution, or a plausible error) that is coherent with the internal knowledge state, the Knowledge Tracing model evolves into a Student Simulator.
    
\begin{table}[ht]
\centering
\small
\resizebox{\textwidth}{!}{%
\begin{tabular}{|l|c|c|c|c|c|c|c|c|}
\hline
\textbf{Model} & \multicolumn{2}{c}{Assist2009} & \multicolumn{2}{c}{Assist2015} & \multicolumn{2}{c}{Junyi} & \multicolumn{2}{c|}{Nips2020} \\
\hline
& AUC & ACC & AUC & ACC & AUC & ACC & AUC & ACC \\
\hline
\multicolumn{9}{|l|}{\textit{DL-based Methods}} \\
\hline
DKT           & 0.7084 & 0.7221 & 0.7093 & 0.7542 & 0.8013 & 0.7200 & 0.7406 & 0.6878 \\
DKVMN         & 0.8157 & -      & 0.7268 & -      & 0.8027 & -      & 0.7673 & 0.7016 \\
SAKT          & 0.8480 & -      & 0.8540 & -      & 0.8340 & 0.7570 & 0.7517 & 0.6879 \\
AKT           & 0.7767 & 0.7532 & 0.7211 & 0.7518 & 0.8948 & 0.8215 & 0.7494 & 0.6930 \\
LPKT          & 0.7788 & 0.7325 & -      & -      & 0.7689 & 0.8344 & -      & -      \\
LBKT$^{\dagger}$ & 0.7863 & 0.7380 & -   & -      & 0.7723 & \textbf{0.8362} & -      & -      \\
AT-DKT        & 0.7574 & 0.7172 & -      & -      & 0.7581 & 0.8325 & 0.7816 & 0.7145 \\
MRT-KT        & 0.8223 & 0.7841 & -      & -      & -      & -      & -      & -      \\
\hline
\multicolumn{9}{|l|}{\textit{PLMs-based Methods}} \\
\hline
BiDKT         & 0.7651 & -      & 0.6766 & -      & -      & -      & -      & -      \\
MLFBK         & 0.8524 & -      & -      & -      & -      & -      & -      & -      \\
LBKT$^{\ddagger}$ & -   & -      & -      & -      & 0.8510 & 0.8320 & -      & -      \\
\hline
\multicolumn{9}{|l|}{\textit{Context-Aware Methods}} \\
\hline
DCL4KT-A      & 0.8153 & -      & -      & -      & -      & -      & -      & -      \\
EERNN         & -      & -      & -      & -      & 0.8370 & 0.7580 & -      & -      \\
EKT           & -      & -      & -      & -      & 0.8420 & 0.7590 & -      & -      \\
RKT           & -      & -      & -      & -      & 0.8600 & 0.7700 & -      & -      \\
\hline
\multicolumn{9}{|l|}{\textit{LLMs-based Methods}} \\
\hline
LLM-FTID      & 0.8393 & 0.7592 & 0.9092 & 0.9092 & 0.8841 & 0.8071 & 0.7890 & 0.6870 \\
LLM-FTTokenID & 0.8143 & 0.7954 & 0.8386 & 0.8813 & 0.8663 & 0.8050 & 0.7774 & 0.5962 \\
LLM-FTText    & 0.8407 & 0.8119 & -      & -      & -      & -      & 0.7762 & 0.7211 \\
GPT-4o        & -      & 0.7274 & -      & -      & -      & -      & -      & 0.6694 \\
\textbf{LLM-KT} & \textbf{0.8870} & \textbf{0.8168} & \textbf{0.9356} & \textbf{0.9185} & \textbf{0.9018} & 0.8294 & \textbf{0.8291} & \textbf{0.7561} \\
\hline
\textit{Imp. (\%)} & +5.68 & +7.59 & +2.90 & +1.02 & +2.00 & +2.76 & +5.08 & +10.06 \\
\hline
\end{tabular}
}
\caption{Comparative results (AUC / ACC) on various KT datasets, as reported in \cite{wang2025llm}.}\label{tab:resultllmkt}
\end{table}

\subsubsection{Using Knowledge Graphs and Heuristics}

The use of KGs and heuristics as structured representations of knowledge has remained an effective strategy for providing agents with contextual grounding. These representations facilitate the simulation of educational profiles and enable more accurate definitions of learners’ knowledge levels.

In \textit{Education in the Era of Neurosymbolic AI}~\cite{jaldi2025education}, the authors propose a symbolic approach to knowledge Modelling by using a manually curated curriculum database structured as a KG. This representation organizes pedagogical concepts and relationships in a format that can be queried by simulated agents or used for assessment tasks. Additionally, the project includes a publicly available interactive repository\footnote{\url{https://github.com/kastle-lab/EduNAILearning-Research}} that supports experimentation and exploration of symbolic educational frameworks.

\begin{figure}
    \centering
    \includegraphics[width=0.7\linewidth]{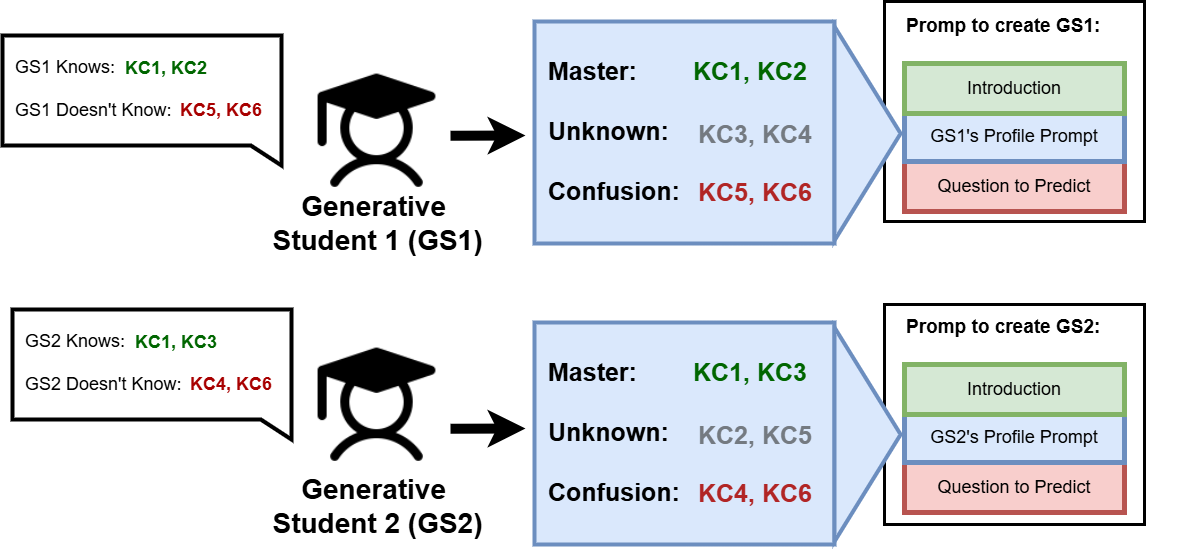}
    \caption{Prompt-based simulation using heuristic-aligned knowledge components.}
    \label{fig:kc_pipeline-label}
\end{figure}

Heuristics, in this context, are treated as KCs and injected directly into LLM prompts to simulate cognitive profiles of students~\cite{lu2024generative}. Each heuristic is accompanied by examples illustrating whether the Simulated Student has mastered, confused, or failed to understand it, as depicted in Figure~\ref{fig:kc_pipeline-label}, which shows how a brief introduction, three color-coded blocks for master (green), unknown (yellow), and confusion (red), and the target question are concatenated to form the final prompt fed to the model. This strategy allows for the generation of responses aligned with varying levels of competence, filtering out content beyond the student’s assumed knowledge scope and analyzing how partial or incorrect understanding affects their responses.

Extending this concept, the KnowLearn system~\cite{haolei2024knowlearn} constructs KGs from real pedagogical data, incorporating feedback, performance history, and contextual learning factors. Unlike static KGs, KnowLearn leverages a Heterogeneous Graph Attention Network (HAN) model to dynamically identify which factors, such as cognitive load or learning intent, most significantly influence student progress. These insights are transformed into personalized inputs embedded into the LLM prompt, enabling the generation of more accurate and adaptive recommendations.

This hybrid approach, combining a knowledge graph generated by a heterogeneous graph attention network (HAN) with a locally deployed LLM, integrates empirically derived insights from graph-based structures to provide personalized learning recommendations based on real student data.

\section{Personality and Individual Traits}\label{sec2}

One of the key factors in the realistic simulation of virtual students is the incorporation of personality traits and individual characteristics. Traditionally, educational models have focused primarily on cognitive variables such as knowledge level or academic performance, often overlooking non-cognitive elements that nevertheless have a direct impact on behaviour and interaction within learning environments. The integration of psychological frameworks such as the Big Five~\cite{mccrae1992introduction} or the Myers-Briggs Type Indicator (MBTI)~\cite{carlson1989affirmative} into generative agents enables the simulation of a more complex and human-like dimension of learning.

This section examines how personality models have been incorporated into LLMs to simulate educational agents with distinct behavioural patterns. It reviews frameworks such as the Big Five for Tutoring Conversation and explores the use of prompting strategies to personalise agent responses based on these profiles. Furthermore, it discusses the pedagogical and practical implications of this approach, including its potential in teacher training and the validation of differentiated instructional strategies.

\subsection{Big Five, BF-TC, and MBTI}

A particularly promising development in the simulation of students using LLMs is the incorporation of psychological models such as the Big Five Personality Traits, also known as the Five Factor Model (FFM), and the MBTI. These models help generate more realistic, diverse, and pedagogically useful student behaviours~\cite{jin2024teachtune, bousalem2025personalized}. The TeachTune study~\cite{jin2024teachtune}, for instance, applies the Big Five model to the design of Pedagogical Conversational Agents (PCAs), enabling the simulation of students with distinct personality-based behaviours. It highlights that personality traits are critical for addressing learner diversity, with teachers consistently identifying personality, motivation, and self-regulated learning strategies as key components of personalised instruction.

The Big Five model is based on five broad personality dimensions: openness to experience, conscientiousness, extraversion, agreeableness, and neuroticism. These traits reflect core aspects of human personality and have a significant influence on learning patterns, motivation, and communication styles~\cite{liu2024personality}. Recent research has shown that certain traits with positive connotations—such as openness, conscientiousness, and emotional stability—can play a key role in educational contexts, although their impact may vary depending on course type and individual learner profiles~\cite{sonlu2025effects}.

Integrating this personality model into LLM-based simulations enables the creation of virtual students with differentiated traits, allowing for varied motivational levels, learning preferences, and behavioural responses. For instance, an agent high in openness may exhibit creativity and curiosity, while one high in conscientiousness is likely to behave in an organised and persistent manner. A notable example is the Big Five for Tutoring Conversation (BF-TC) model, which does not redefine the Big Five traits but rather reformulates their item definitions to suit the context of educational dialogues. This adaptation supports more natural and pedagogically coherent interactions between students and tutoring systems~\cite{liu2024personality} (see Table~\ref{tab:bigfive_bftc}).

In contrast, the MBTI model classifies individuals into 16 personality types based on four dichotomies: extraversion vs. introversion, sensing vs. intuition, thinking vs. feeling, and judging vs. perceiving~\cite{wang-etal-2024-incharacter}. Although more structured and less commonly used in Simulated Student modelling than the Big Five, MBTI has been applied to identify learning styles and improve group dynamics. However, it is important to note that the MBTI remains a controversial instrument within the psychological sciences~\cite{Stein2019Evaluating}, lacking strong empirical support, with concerns regarding its reliability and predictive validity. Many researchers point out that MBTI's type-based approach oversimplifies personality, relies on questionable theoretical foundations, and fails to capture the complexity and fluidity of human traits. Despite these limitations, MBTI has occasionally been integrated into LLM-based simulations alongside the Big Five model, as explored in a recent preprint~\cite{li2025exploring}, where MBTI serves as a broad typological descriptor to enrich agent characterization, while the Big Five provides granular behavioural dimensions—such as conscientiousness, emotional stability, and openness—enabling a more nuanced assessment of profile-behaviour alignment in Simulated Student agents.

Despite their differing theoretical foundations and levels of empirical support, both models are used in this study to construct psychological profiles of virtual students~\cite{li2025exploring}. This approach enables the simulation of classroom scenarios involving students with diverse behaviours and learning challenges, offering a valuable opportunity to refine pedagogical strategies and promote metacognitive development without ethical risks. Tools such as SENEM-AI and TutorUp employ LLM-generated student agents with varied psychological traits to help prospective educators recognize and respond to a wide range of classroom situations~\cite{pan2025tutorup, pentangelo5165658senem}. These simulations facilitate the practice of differentiated motivational techniques, the delivery of personalized feedback, and the exploration of inclusive instructional methods. However, the psychological coherence and educational validity of these simulated profiles require closer scrutiny, especially when based on contested typologies such as the MBTI.

\begin{table}[ht]
\renewcommand{\arraystretch}{1.5}
\centering
\resizebox{\textwidth}{!}{%
\begin{tabular}{|p{3.7cm}|p{5.8cm}|p{7cm}|}
\hline
\textbf{Trait} & \textbf{Big Five (General)} & \textbf{BF-TC (Tutoring Adaptation)} \\
\hline
\textbf{Openness} & Curious, imaginative, open to new ideas and experiences & Creativity in answers, curiosity, openness to teacher suggestions \\
\hline
\textbf{Conscientiousness} & organised, disciplined, reliable & Well-organised responses, effort in language learning, task focus \\
\hline
\textbf{Extraversion} & Sociable, talkative, outgoing & Active participation in conversation, talkative, confident in speaking \\
\hline
\textbf{Agreeableness} & Compassionate, cooperative, empathetic & Polite, empathetic, cooperative attitude during tutoring \\
\hline
\textbf{Neuroticism} & Anxious, insecure, emotionally unstable & Nervous during conversation, hesitant, worries about being wrong \\
\hline
\end{tabular}%
}
\caption{Comparison between standard Big Five traits and BF-TC adaptation for tutoring dialogues.}
\label{tab:bigfive_bftc}
\end{table}

To ensure the quality and fidelity of these simulations, researchers have introduced multi-aspect validation frameworks that combine expert annotation with automated LLM-based scoring. These frameworks compare the generated personality profiles with standardized instruments such as the Big Five Inventory (BFI)~\cite{liu2024personality}. The BFI, a well-validated measure of global personality traits, is used in Liu \emph{et al.} to test whether the Big Five for Tutoring Conversation (BF-TC) model accurately encodes personality information within educational dialogues. Their results show a strong correspondence between the BF-TC labels and the BFI categories, indicating that the adapted model captures conversational expressions of personality traits effectively.

Further research using the abbreviated Big Five Inventory-2-XS (BFI-2-XS)~\cite{sonlu2025effects} has assessed the perceived personality of learning agents, both embodied (with physical representation) and disembodied. Results show generally positive evaluations across all traits except neuroticism, with agents exhibiting higher levels of positive traits receiving the most favorable ratings.

\subsection{Prompting Techniques}

Various prompting techniques have been employed to guide LLMs in simulating student behaviours, knowledge levels, and responses within educational contexts. These techniques aim to generate realistic and informative simulations that support applications such as intelligent tutoring system evaluation, question quality assessment, and the understanding of learning processes.

A widely adopted approach frames the LLM not as the learner but as an instructor that predicts a student’s response \citep{lu2024generative}. This pedagogical stance yields predictions that align more closely with actual student profiles. Rather than merely labeling a learner’s proficiency, prompts that embed illustrative examples of both mastered and misunderstood rules capture student understanding more faithfully. For example, a prompt can include correct answers for mastered rules and incorrect answers for confused rules to simulate misconceptions.

Prompt engineering under this approach follows a structured template. It begins with a task description, continues with a detailed student profile that lists known, confused, and unknown rules, and ends with a multiple-choice question to be answered by the Simulated Student. By systematically varying the rule sets, researchers can generate a broad spectrum of student profiles.

To further enhance realism, several strategies have been proposed to introduce uncertainty and variability into the simulations. These include focal confusion prompting, explicitly Modelling ``unknown'' knowledge components, and presenting students with questions of varying difficulty~\cite{lu2024generative}. Few-shot prompting, which involves providing several input-output pairs to guide the LLM, is another explored method. Though not always central, it remains a general strategy applicable in ITS or even student simulation scenarios~\cite{garcia2024review}

Chain-of-Thought (CoT) prompting, which incorporates explicit reasoning steps to improve model performance (as shown in a recent preprint~\cite{zhang2024spl}), is also discussed as a way to enhance human-like responses and adaptability. However, recent findings in the context of simulating student answers to exam questions indicate no direct correlation between the quality of CoT-generated rationales and the accuracy of simulated responses~\cite{benedetto2024using}. This suggests that while CoT can be employed to improve interpretability, it does not necessarily yield more accurate student-level simulations. Nevertheless, CoT prompting can be purposefully used to simulate flawed reasoning, where the model articulates step-by-step thought processes that include typical misconceptions or calculation errors. This adds educational value by making visible how a student might arrive at an incorrect answer.

Including expressions of uncertainty such as “I think” or “I’m not sure” can enhance the naturalness and believability of Simulated Student responses, especially when compared to overly direct or repetitive answers. However, the reviewed works do not systematically analyze the use of hesitation markers, such as “uh” or “hmm”, nor do they provide empirical evidence that these elements contribute to more credible simulations. 
Similarly, there is no documentation of Simulated Students posing metacognitive follow-up questions such as “Is that what you meant?”, as follow-up initiatives are typically attributed to the tutor side of the interaction . Therefore, while hedging appears to support more human-like and appropriately uncertain student behaviour, the contribution of hesitation markers and self-reflective prompts remains largely unexplored~\cite{jin2024teachtune, xu2025classroom, 10937459}.

Simulated profiles can also integrate personality traits and learning styles to further increase realism. Leveraging psychological models such as the Big Five or MBTI, prompts can specify attributes like introversion, openness, or conscientiousness, influencing the tone and structure of student responses. For instance, a highly extroverted student might respond enthusiastically and at length, while a more cautious or anxious student may provide shorter, tentative answers. 
Prompt formulations that specify emotional reactions to difficulty, such as frustration, excitement, or insecurity, enable the simulation of affective states, thereby enhancing the authenticity of simulated dialogues~\cite{chen2024persona, chu2025llm, xu2024eduagent, liu2024personality}.
In addition, where the teacher-agent must model the student-agent's perspective, where the LLM is instructed to act as a teacher imagining how a student might respond, can help constrain the model’s knowledge to the Simulated Student’s profile. This framing helps avoid the leakage of expert knowledge and better aligns the response with the expected limitations of the student. This technique has shown promise in achieving more realistic performance aligned with different skill levels~\cite{lu2024generative}.

Prompt engineering can benefit from a structured template that includes context, task instructions, response format guidelines, and illustrative examples when applicable. This structure not only improves clarity for the model but also ensures consistency across different simulation tasks. Varying the formulation of these prompts or adjusting parameters such as response length or temperature can introduce desirable variability, preventing the generation of repetitive or overly scripted outputs.

In Personality-aware Student Simulation for Conversational Intelligent Tutoring Systems~\cite{liu2024personality}, a framework is proposed that leverages LLMs for personality-aware student simulation in a language learning context. While the presented examples suggest tutor-led conversational interactions with predefined scaffolding categories, the study specifies that pedagogical instructions are directly embedded within the prompts assigned to the tutor role. Furthermore, student responses are generated by a simulator that, although conceptually distinct, relies on the same type of LLM as the tutor to perform its role. The overarching goal is to enhance human-machine interaction and support the development of conversational ITS in domains such as language learning.

\begin{figure}
    \centering
\includegraphics[width=1\linewidth]{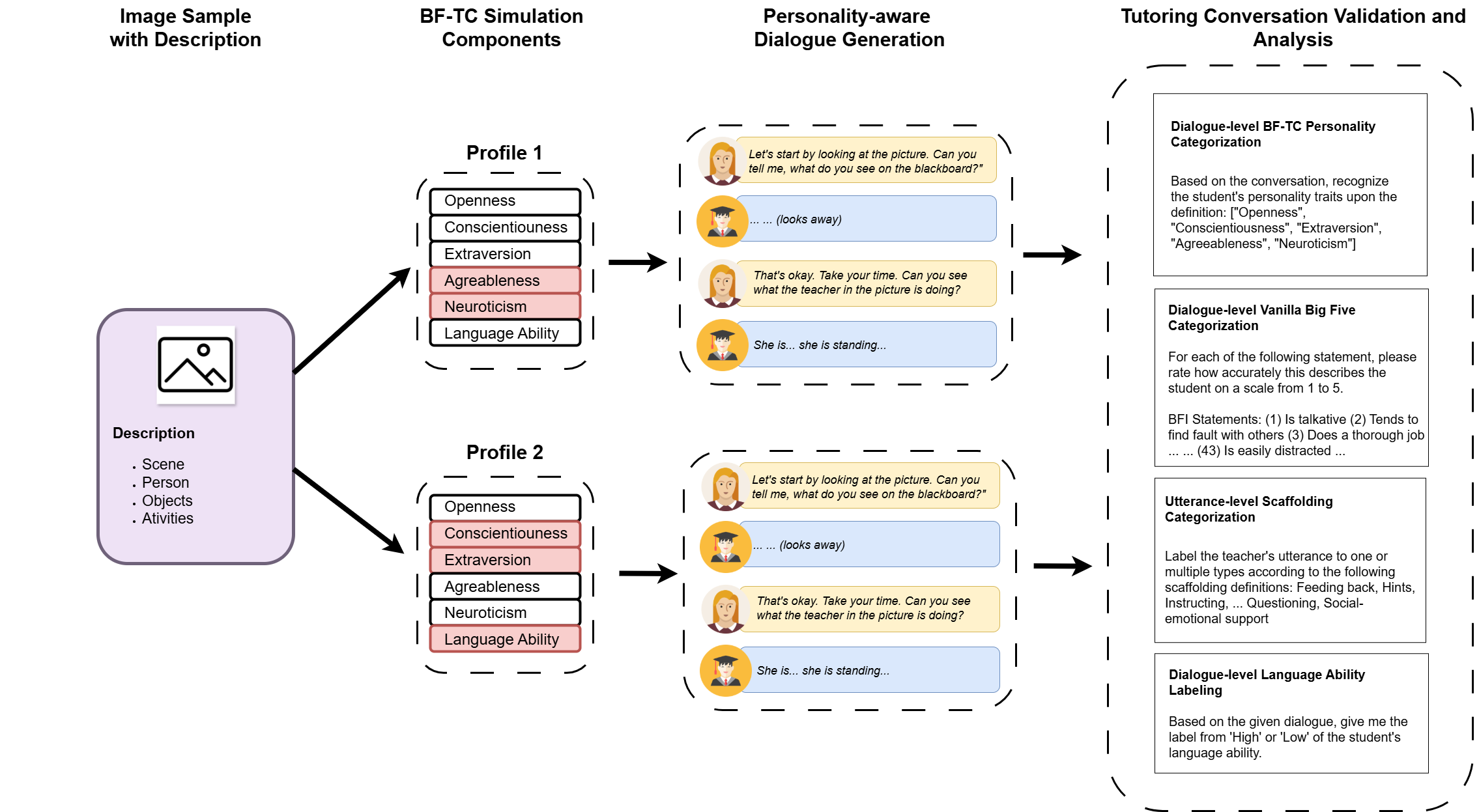}
    \caption{Framework for personality-aware student simulation with LLMs, enabling personalized dialogue generation and multi-aspect validation.}
    \label{fig:BFTC}
\end{figure}

Figure~\ref{fig:BFTC} illustrates the complete simulation pipeline. The process begins with a visual input that establishes the context for the dialogue. The BF-TC simulation engine assigns each Simulated Student a combination of personality traits, such as extraversion and neuroticism, and a corresponding language ability level.
These attributes then condition the student's behaviour in a dialogue generated by the LLM. The figure also outlines the subsequent multi-aspect validation process, which includes: (1) personality categorization based on BF-TC definitions, (2) standard Big Five Inventory (BFI) assessments, (3) classification of language ability, and (4) fine-grained labeling of the tutor’s scaffolding strategies. This integrated approach ensures that Simulated Student behaviour is both coherent and pedagogically meaningful.

Prompting strategies can be designed to guide LLMs in generating pedagogically sound and emotionally attuned feedback in counselor education. In the context of intelligent tutoring systems, prompts are constructed to reflect established feedback frameworks such as the Sandwich, WWW, and STATE methods~\cite{rudolph2024automated}. The Sandwich method frames constructive critique between two positive observations to balance encouragement and correction. The WWW approach structures feedback into perception (I noticed that...''), impact (This made me feel...''), and wish (``I would appreciate if in the future...''), promoting clarity, empathy, and actionable guidance. The STATE method further deepens this by encouraging dialogue, including the counselor’s perspective, and phrasing feedback tentatively to foster reflection and openness. These prompting techniques not only shape the tone and structure of LLM-generated responses, but also increase the alignment of feedback with instructional goals and learner acceptance.

Finally, simulating student intention has also been explored in Intelligent Tutoring Systems. In this context, LLMs are fine-tuned on datasets specifically annotated for educational intent classification, such as LAAIIntentD~\cite{keerthichandra2024large}. Although prompt engineering details during inference are not fully disclosed, the fine-tuning process involves presenting diverse student interaction scenarios to improve recognition accuracy.

To maximise the educational utility of these simulations, it is crucial to balance realism with coherence, ensuring that student profiles are maintained consistently across multi-turn interactions while allowing for natural variability. Simulated Students should not appear overly robotic or uniformly skilled across all topics. Instead, introducing controlled uncertainty, emotional expression, and context-aware behaviour helps create more believable and pedagogically valuable student simulations.

\subsection{Role-Playing and Alternative Simulations}

An important area within LLM-based simulation involves agents that assume roles beyond the traditional student–teacher dynamic. These agents are often designed to portray fictional or historical characters, enabling the creation of immersive, interactive environments that are rich in nuance. Advances in this area can directly benefit student simulation by informing more sophisticated, engaging, and context-aware educational agents. This section reviews several contemporary approaches that highlight the implementation and evaluation of role-playing agents in learning contexts.

One of the most prominent approaches in historical simulation is the Character-LLMs framework, where models such as LLaMA 7B are fine-tuned to simulate specific historical figures through a process called Experience Upload~\cite{shao2023character}. In this method, curated biographical profiles (e.g., Beethoven, Cleopatra) are converted into narrative scenes that include interactions, emotions, and internal thoughts. These scenes, combined with a small set of “protective experiences” designed to prevent anachronisms or inconsistencies, are used in a supervised fine-tuning process, with individual models trained on 1,000 to 2,000 scenes per character. Evaluation is conducted via simulated interviews (single-turn and multi-turn), and responses are assessed by a GPT-3.5-based automatic judge using five criteria: personality, memorization, consistency, values and hallucinations. Results show that these character-specific models outperform instructional alternatives like Alpaca or Vicuna in character fidelity, in some aspects rivaling even ChatGPT despite being smaller in scale. This approach is complemented by platforms like HistoryLens, an educational tool where students interact with historical figures via role-play dialogues. This interaction fosters deeper contextual understanding and supports the development of historical empathy~\cite{breen2025large}. Together, these initiatives highlight the pedagogical potential of historically grounded role-play simulations for immersive and personalized learning.

Building on the goal of aligning agent responses with role identity, another significant contribution is Role-Conditioned Instruction Tuning (RoCIT)~\cite{wang2023rolellm}. This method fine-tunes models such as LLaMA and ChatGLM2 using two sources: stylized responses generated via RoleGPT and contextual QA triples from Context-Instruct. These are combined with system instructions specifying role name, description, typical phrases, and tasks. This methodology eliminates the need for heavy context-retrieval mechanisms, enabling more direct personalization. Evaluation metrics include Rouge-L, automatic scoring via GPT-4, and human annotation of style, accuracy, and knowledge across three data conditions: no role (RAW), stylized (CUS), and knowledge-enhanced (SPE).

In parallel with these technically grounded approaches, recent work has explored the narrative coherence of role-playing agents. From a storytelling perspective, TIMECHARA introduces the notion of point-in-time hallucinations, evaluating whether a character maintains temporal consistency within its fictional universe~\cite{ahn2024timechara}. The NARRATIVE-EXPERTS method enhances consistency for characters such as Harry Potter or Frodo by segmenting reasoning into space–time stages. Meanwhile, introduced in a preprint, the ERABAL framework applies boundary-aware learning to assess role fidelity by presenting questions that challenge the agent's knowledge limits~\cite{tang2024erabal}.

As a complement to these narrative techniques, a preprint entitled \textit{The Oscars of AI theater: A study on role-playing with language models}~\cite{chen2024oscars} proposes a broader structural taxonomy. This work categorizes LLM-based role-playing agents into real-world personas and fictional characters, introducing a comprehensive evaluation framework that combines human judgment, reference-based scoring, and automated metrics. Additionally, the conceptual framework \textit{From persona to personalization: A survey on role-playing language agents}~\cite{chen2024persona} categorizes agents into demographic, character-based, and individualized types, emphasising the value of hybrid evaluation methods that assess fidelity, coherence, and empathy.

Translating these role-based strategies into practical applications, recent studies have begun to explore social and educational contexts. For instance, simulations of civic debates between fictional citizens have been used in local government workshops, with GPT-4 acting as facilitator, evaluator, and ideation engine~\cite{sato2024llm}.

Recent advancements include efforts to simulate comprehensive interactive environments. A recent preprint~\cite{yu2024affordablegenerativeagents} investigates the Affordable Generative Agents (AGA) framework in virtual settings such as those found in the Generative Agents framework and VirtualHome. Within these simulations, generative agents are designed to develop routines through learned policies, cultivate social relationships by Modelling inter-agent interactions, and respond to external environmental changes. The evaluation of this framework employs established methods, including simulated interviews to assess memory and future planning, and automated assessments, leveraging models like GPT-4 as evaluators, to gauge behavioural coherence and reactivity. This rigorous approach aims to enhance the believability and practical applicability of LLM-driven interactive agent systems.

Despite the growing sophistication of role-playing agents, important limitations remain underexplored in current implementations. Many systems risk reducing complex historical or cultural identities to tokenised caricatures or folklorised narratives, especially when agents are built from simplified biographical or demographic templates~\cite{chen2024persona}. This raises ethical and pedagogical concerns regarding stereotype reinforcement, cultural oversimplification, and authenticity. Moreover, while existing evaluation frameworks assess character fidelity, narrative coherence, and stylistic appropriateness, few address the educational impact of these simulations. A distinction must be made between stylistic credibility, how convincing an agent sounds, and pedagogical efficacy, whether the interaction facilitates meaningful learning outcomes. Future research should therefore prioritise hybrid evaluations that align expressive realism with instructional value, while remaining sensitive to the sociocultural implications of simulating identity.

\section{Evaluation}

Assessing the quality and utility of Simulated Students is essential to validate their application in educational contexts. Beyond verifying correct answers, effective evaluation must consider the extent to which these agents replicate human-like cognitive processes, behaviours, and individual traits. Current methodologies combine automated metrics, human judgement, and pedagogical or psychometric frameworks to assess fidelity, usefulness, and adaptability across learning scenarios.

A key aspect of this process involves analysing the datasets used to train and evaluate these agents. Publicly available resources, ranging from behavioural logs and knowledge tracing corpora to annotated dialogues and synthetic student models, shape the development and validation of simulations. Their structure, scope, and representativeness influence both performance and generalizability. The following subsection reviews prominent datasets, highlighting their role in supporting more robust, multimodal, and context-aware educational simulations.

\subsection{Datasets Resources}

MoocRadar~\cite{yu2023moocradar} provides a multi-aspect repository comprising exercises, conceptual mappings, and behavioural interaction records. It is particularly useful for cognitive modelling tasks and holds considerable potential for structured knowledge-based student simulation.

Publicly available datasets play a crucial role in training and evaluating simulations involving LLMs. Among the most prominent is the EduAgent~\cite{xu2024eduagent} framework, which is distributed in two main versions:
\begin{itemize}
    \item \textbf{EduAgent310} (N=310): captures detailed real-world behavioural data during short classroom sessions, including eye gaze, mouse movements, and inferred cognitive states.
    \item \textbf{EduAgent705} (N=705): consists of synthetic student agents generated with individualised attributes such as attitude, concentration level, and prior knowledge.
\end{itemize}

The EduAgent dataset, which includes behavioural and learning data, is commonly used to validate LLM-based student simulations. It reflects a broader trend toward developing rich, multimodal resources for more nuanced Modelling of student behaviour. However, while a variety of other datasets and simulation frameworks have been developed to support the training, evaluation, and refinement of these agents, most lack access to the actual lecture content or subject matter, and typically omit pre-topic assessments, providing only final questionnaires.

These resources vary significantly in scale, structure, and purpose, ranging from detailed interaction logs to synthetic corpora and annotated dialogue datasets, highlighting both the progress made and the remaining gaps in the field.

Several datasets have been specifically developed for knowledge tracing, including ASSIST09, ASSIST12, and Junyi7\citep{fu2024sinkt}. Although these datasets expand the student sample size (2,661, 2,000, 2,000), they do not provide the actual text of the questions, relying solely on question identifiers. This limitation prevents them from addressing key shortcomings previously identified in Agentedu. In contrast, the MoocRadar corpus~\cite{yu2023moocradar} distinguishes itself by its scale and annotation granularity, comprising 2,513 fully documented exercises, 5,600 fine-grained knowledge concepts, and over 12 million recorded interactions, all accompanied by open-source tools publicly available on GitHub.

Evaluation-focused datasets have also emerged. The preprint TRUSTSIM benchmark~\cite{huang2024social} introduces 740 scenarios aimed at assessing the reliability of social simulations, not limited to student Modelling, across 10 disciplines using 14 different LLMs. The authors further propose the AdaORPO method to enhance response consistency in such contexts. Additional studies leverage corpora such as ARC, RACE, and CUP\&A~\cite{benedetto2024using} to evaluate whether the quality of simulated answers varies according to the assigned student level in multiple-choice questions (MCQs).

Efforts to annotate dialogue acts have leveraged Active Learning strategies within the Data Maps framework. One annotated corpus comprises 2,156 tutor utterances and 1,470 student responses, covering 31 dialogue acts. The CoreMSE technique is reported to reduce data requirements by approximately 30\%~\cite{tan2023does}.

\begin{table}[p]
  \centering
  \vspace{1em} 
  \rotatebox{90}{%
    \begin{adjustbox}{
        max width=\dimexpr\textheight-1.5cm\relax,
        max height=\dimexpr\textwidth-1cm\relax
      }
      \small
      \begin{tabular}{|p{3cm}|p{2.5cm}|p{3.5cm}|p{1.9cm}|p{3.5cm}|p{3.5cm}|p{3cm}|p{3cm}|}
        \hline
        \textbf{Dataset} & \textbf{Domain} & \textbf{Educational Level} & \textbf{Assessment Design} & \textbf{Key Measured Variables} & \textbf{Data Volume} & \textbf{Content Coverage} & \textbf{Assessment Item Content} \\
        \hline
        OULAD~\cite{kuzilek2017open} & Online HE, STEM and Social Sciences & Higher Education & No, formative assessments only & Assessment scores, VLE activity logs & 32.593 students, 22 courses & Not available & Not available \\
        \hline
        MoocRadar~\cite{yu2023moocradar} & MOOCs, multi-disciplinary & Adult/Continuing Education & No, exercises only & Exercise correctness, Bloom level & 14.224 students, 2.513 exercises, 5.600 concepts & Concept labels only & Textual content and question types \\
        \hline
        TrustSun~\cite{huang2024social} & Computational Social Science & Mixed (simulated profiles) & Yes, pre/post & Self-report consistency, scenario engagement & 740 instances across 10 domains & Scenario text included & Evaluation prompts included \\
        \hline
        LPCA~\cite{ndihokubwayo2022dataset} & Physics, Light Phenomena & Secondary Education & Yes, pre/post & 30-item conceptual test & 251 students & Light and optics topics & Multiple-choice plus one open response \\
        \hline
        GOCUT~\cite{ndihokubwayo2022dataset} & Physics, Geometric Optics & Secondary Education & Yes, pre/post & 25-item conceptual test & 136 students & Conceptual understanding of geometric optics & Multiple-choice plus drawings \\
        \hline
        QPCS \& QPAT~\cite{nyirahabimana2024university} & Physics, Quantum & Higher Education & Yes, pre/post & Conceptual test (32 items) and attitude survey (24-point Likert) & 385 students & Introductory quantum physics & Full QPCS and QPAT instruments included \\
        \hline
        EduData~\cite{gao2025agent4edu} & Math and Physics, China & Secondary Education & No & Correct/incorrect responses & 18.045 logs, 500 students, 458 concepts & Concept tags only & Text not included \\
        \hline
        
        ASSISTments09~\cite{10.1007/s11257-009-9063-7} & Mathematics, ITS & Secondary Education & Yes, item-level & First-attempt accuracy, attempts, timing & 4.217 students, 346.860 logs & K–12 math concepts (KC IDs) & Exercise IDs only \\
        \hline
        edX HarvardX-MITx~\cite{HarvardX} & MOOCs, 16 courses & Adult/Continuing Education & No & Enrollment, activity, certification & 641.138 users & Performance and activity data by course & Aggregated metrics only \\
        \hline
        OLI Statics F2011~\cite{zhang2017incorporating} & Engineering, Statics & Higher Education & No & Step correctness, hint use, timing & 332 students, 257k record & Mechanics (statics) content & Exercise labels, descriptions of numbered knowledge components, and problem-level features. \\
        \hline
        SocraticMATH~\cite{ding2024boosting} & Mathematics, dialogues & Primary Education & N/A, dialogue-based & Dialogue quality (human and automated) & 6 846 dialogues, 513 concepts & Arithmetic, factors, divisibility & Socratic dialogues with embedded Q/A \\
        \hline
        University Students – Quantum (Rwanda)~\cite{nyirahabimana2024university} & Physics, Quantum & Higher Education & Yes, pre/post & QPCS scores, attitudes, classroom observations & 385 students & Introductory physics course on quantum & Full test and survey instruments \\
        \hline
        KDD Cup 2010~\cite{yu2010feature} & Math, Algebra & Secondary Education & No & First-attempt accuracy, step tracking & > 10.000 students, > 30 M step logs & Algebra skills (pre-university) & Step IDs only \\
        \hline
        EduAgent 310/705~\cite{xu2024eduagent} & Short ML lecture & Diverse educational backgrounds & Yes, post only & Gaze, mouse, cognitive states, quiz accuracy & 310 real plus 705 agents & ML content via slides & MCQ quiz items \\
        \hline
        EdNet~\cite{choi2020ednet} & English, TOEIC prep & Adult/Continuing Education & No & Correctness per attempt, sequence logs & 131 M logs, 784.309 users & TOEIC-aligned learning units & Item IDs only \\
        \hline
      \end{tabular}
    \end{adjustbox}%
  }
  \vspace{1em} 
  \caption{Summary of educational datasets with unified attributes and reference citations.}
  \label{tab:datasets}
\end{table}

The SPOCK system~\cite{sonkar2023class}, an LLM adapted for tutoring tasks, is trained on two public corpora: a scaffolding dataset with 648 problems and sub-problems, and a conversational set with 648 dialogues. Complementarily, the 3DG framework~\cite{zhang20243dg} generates synthetic dense datasets using GANs and GPT-4 based on binary performance data from 118 adult learners.

To better understand the diversity and scope of existing resources used in the simulation of students and educational environments, Table~\ref{tab:datasets} summarizes representative datasets frequently employed in recent research. These datasets vary in domain, structure, and data type, covering areas such as tutoring dialogues, knowledge tracing, behavioural Modelling, and social simulation. The table highlights key characteristics, including the presence of textual content, evaluation mechanisms, measurement types, and the incorporation of sensor-based data such as eye tracking and mouse movement.

\subsection{Human Evaluation Methods}

Human evaluation remains a critical method for validating the realism, pedagogical utility, and coherence of Simulated Student agents generated by LLMs. Unlike automated metrics, human-centered strategies can assess nuanced dimensions such as instructional relevance, emotional realism, and social believability, factors essential in educational simulations.

Human evaluation of LLM-based educational agents extends beyond Simulated Students and includes intelligent tutoring systems, which share methodological similarities through the simulation of instructional roles. In both cases, human evaluators assess the pedagogical coherence, relevance, and realism of AI-generated interactions within educational settings. Several studies have incorporated domain experts into this evaluation process. Sonkar et al.~\cite{sonkar2023class} involved postgraduate biology experts to assess the quality of scaffolding datasets generated by LLMs, focusing on factual accuracy, contextual relevance, and instructional usefulness. Similarly, Roest et al.~\cite{roest2024next} conducted two complementary studies involving both students and teachers to compare AI-generated hints with traditional forms of support, highlighting the perceived efficacy and pedagogical value of LLM-driven guidance.

Teacher training environments have also adopted Simulated Students as evaluation targets, a trend highlighted in reviews of Affective Intelligent Tutoring Systems. For instance, the scoping review by Fernández-Herrero et al.~\cite{fernandez2024evaluating} discusses studies where systems, such as SimInClass, utilize quasi-experimental designs involving pre-service teachers to assess their adaptation to AI-mediated instruction and changes in emotional states. Dai et al.~\cite{dai2024designing} complemented this with a qualitative case study in VR, triangulating data through think-aloud protocols, interviews, and system logs to evaluate the realism and engagement of AI-based agents.

Broader benchmarks such as MRBench~\cite{maurya2024unifying} have formalized human validation protocols at scale MRBench focuses on pedagogical feedback to student misconceptions, using expert judgment across eight instructional dimensions and reporting high inter-rater reliability (Fleiss’ $\kappa = 0.65$). Both benchmarks reinforce the value of structured human evaluations aligned with instructional goals.

A more direct approach to validating student agents is presented by Jin et al.~\cite{jin2024teachtune}. They developed a structured pipeline that allows educators to configure Simulated Students by defining their knowledge levels and various learning traits such as self-efficacy and motivation. Teachers evaluated the realism of these agents through technical evaluations involving blind prediction tasks and interactions with chat transcripts. The evaluations showed low error rates, specifically a 5\% median error for knowledge components and a 1.3 median difference (out of 12) for student traits, indicating that the Simulated Students' behaviours closely aligned with teachers' expectations. A user study with 30 educators further demonstrated that automated chats with these simulations significantly helped teachers explore a broader range of student traits and reduced their perceived task load.

The document introduces PersonaEval~\cite{zhangpersonaeval} , a benchmark designed to assess the effectiveness of LLMs in role-playing evaluation tasks. This benchmark is based on real-world data from the wired 5 Levels video series, where experts explain concepts to five distinct audiences (a child, a teenager, an undergraduate student, a graduate student, and another expert). The objective is to evaluate the ability of LLMs to classify and distinguish specific roles based solely on linguistic cues.

Collectively, these studies demonstrate that human evaluation provides unique insights into the realism, usefulness, and pedagogical alignment of Simulated Student agents. Nonetheless, limitations persist—including evaluator subjectivity, limited scalability, and the absence of standardized protocols. Future work should explore hybrid approaches that integrate human judgment with LLM-based assistance to enable scalable, credible, and context-sensitive validation of educational simulations.

\subsection{Automatic Evaluation Methods}

While the previous section reviewed human evaluation, validating student simulations at scale requires reliable automatic methods. The studies in this section, while often focused on assessing real student work, provide the foundational methodologies and tools used to automatically evaluate the outputs (e.g., text, code, interactions) generated by Simulated Students. This integration of LLMs and other algorithms for automatic assessment represents a significant trend in educational technology, with a growing body of research exploring how artificial intelligence can be leveraged to deliver scalable, efficient, and pedagogically sound evaluations, thereby assessing the simulations' fidelity and performance.

One notable application involves the use of LLMs as evaluators of student-generated texts, such as responses to questions or essays. For example, \textit{Automatic Assessment of Student Answers using Large Language Models: Decoding Didactic Concepts}~\cite{schonle2024automatic} examines machine learning techniques, including DistilBERT, for automating text-based assessment in virtual learning environments. This study employs Transformer-based data augmentation combined with grammar-enhanced feature selection, including techniques such as part-of-speech analysis. While traditional models like SVMs initially outperformed DistilBERT on synthetic datasets, the integration of linguistic features significantly enhanced DistilBERT’s performance, highlighting its potential for processing complex language structures in automated evaluation tasks.

In programming education, LLMs have proven effective in assessing code-based exercises. \textit{Feedback-Generation for Programming Exercises With GPT-4}~\cite{azaiz2024feedback} investigates GPT-4 Turbo's capability to generate formative feedback based on student-submitted code. The findings demonstrate GPT-4's proficiency in interpreting programming solutions and providing targeted, constructive feedback in authentic learning scenarios.

The evaluative capabilities of LLMs have also been explored in role-based interaction contexts. \textit{PersonaEval: Benchmarking LLMs on Role-Playing Evaluation Tasks}~\cite{zhangpersonaeval} introduces a benchmark to assess LLM effectiveness in role-play classification tasks. By framing the problem as a classification challenge, the study investigates whether LLMs can distinguish statements from personas with varying levels of expertise, suggesting strong potential for LLMs to assess Simulated Students in complex, interactive environments.

LLMs and related algorithms also contribute to other simulation-based educational tasks. In a recent preprint titled \textit{QG-SMS: Enhancing Test Item Analysis via Student Modelling and Simulation}~\cite{nguyen2025qg}, student simulation is employed to improve test item analysis. While the primary focus is on question generation, the framework’s underlying student models facilitate automated evaluation of simulated learner performance on assessment items.

LLMs are also increasingly utilised to verify the consistency and accuracy of feedback provided by intelligent tutoring systems. The work \textit{Large Language Models as Evaluators in Education: Verification of Feedback Consistency and Accuracy}~\cite{seo2025large} highlights this emerging role of LLMs as quality control agents, emphasising the importance of reliable feedback in simulations involving virtual students.

Several reviews support and contextualize these findings. For instance, a systematic review by González-Calatayud et al., referenced in~\cite{schonle2024automatic}, reveals a prevailing emphasis on formative assessment and automated grading within AI-driven evaluation systems. This reflects a broader trend toward integrated and scalable educational assessment supported by AI.

Similarly, \textit{Evaluating Recent Advances in Affective Intelligent Tutoring Systems: A Scoping Review}~\cite{fernandez2024evaluating} considers the role of automated assessment in affective tutoring systems (ATSs). While the primary focus is on emotional state evaluation, the potential integration of LLMs is suggested to enhance interactions and facilitate a shift toward more comprehensive, multi-dimensional evaluation within intelligent systems.

Finally, \textit{Empowering Education by Developing and Evaluating Generative AI-Powered Tutoring System for Enhanced Student Learning}~\cite{banjade2024empowering} describes a system where course materials are reviewed by a tutor before being delivered to students. This review process may include AI-based assessment mechanisms, underscoring the importance of such technologies in ensuring the pedagogical integrity of learning platforms that incorporate Simulated Student interactions. As shown in Table~\ref{tab:evaluation_comparison}, human and automatic evaluation methods offer distinct strengths and limitations when assessing the behaviour and effectiveness of a Simulated Student.

\begin{table}[h!]
\centering
\resizebox{\textwidth}{!}{%

\begin{tabular}{|p{2.5cm}|p{6cm}|p{6cm}|}
\hline
\textbf{Criterion} & \textbf{Human Evaluation} & \textbf{Automatic Evaluation} \\
\hline
\textbf{Key \newline Participants} & Students, educators, domain experts, researchers & LLMs and other machine learning models, including DistilBERT, GPT-4, and automated classification algorithms \\
\hline
\textbf{Evaluation Types} & Qualitative feedback, surveys, expert judgments, comparisons with real-world data & Automated answer scoring, consistency verification, performance analysis, semantic classification \\
\hline
\textbf{Advantages} & Contextual understanding, pedagogical insight, emotional interpretation, expert-driven assessments & Scalability, speed, consistency, ability to process large-scale data \\
\hline
\textbf{Limitations} & Subjectivity, variability among evaluators, lack of standardization, high resource demands & Limited pedagogical nuance, potential training biases, dependency on data quality and prompt engineering \\
\hline
\textbf{Representation of Diversity} & Includes certain underrepresented groups, such as Deaf learners, though representation remains limited overall & Frequently fails to authentically capture learner diversity, often portraying stereotypical or idealized behaviours \\
\hline
\textbf{Notable \newline Applications} & Evaluation of conversational agents, instructional feedback analysis, teacher training simulations & Assessment of student-generated text, code evaluation, test item analysis, role-based interaction assessment \\
\hline
\textbf{Primary \newline Objective} & To validate realism, pedagogical relevance, and instructional effectiveness in simulated educational contexts & To measure correctness, consistency, technical quality, and alignment with instructional goals \\
\hline
\end{tabular}
}
\caption{Comparative analysis of human and automatic evaluation methods in LLM-Based educational simulations.}
\label{tab:evaluation_comparison}
\end{table}

\subsection{Evaluation Metrics Employed}

A wide range of evaluation metrics have been employed across studies to assess the performance and reliability of systems using LLMs in educational contexts. Among the most commonly used is accuracy, often applied to specific tasks involving LLMs under fine-tuning conditions with custom input data to evaluate model responses, as demonstrated in prior work~\cite{schonle2024automatic, xu2025classroom}. Table~\ref{tab:evaluation_metrics} summarizes these metrics, categorising them by focus area, providing specific examples, and outlining their intended purpose within educational evaluation.

The utility of AI-generated questions has been assessed through both automatic and human-centered approaches. For instance, Nguyen et al.~\cite{bhat2022towards} used the T5 model alongside an information score metric and manual annotations from human raters. Their study also compared GPT-3’s assessments of question utility with human evaluations, examining inter-rater agreement.

Regarding the internal consistency and perceived relevance of AI-generated questions, the review by Garc{\'\i}a-M{\'e}ndez~\cite{garcia2024review}  highlights the work of Nasution~\cite{nasution2023using}, who applied Cronbach's alpha to assess their reliability. Complementarily, Nasution also utilised student surveys to measure subjective aspects such as clarity, relevance, and depth.

Within the programming domain, syntactic and functional correctness of GPT-4-generated feedback was validated via manual inspection and unit testing, using precision, recall, and accuracy as core metrics~\cite{azaiz2024feedback}. Other studies evaluated semantic similarity using BERT-based methods and information entropy to assess the richness and relevance of generated scaffolding, comparing it to baseline outputs from GPT-3.5~\cite{10.1007/978-3-031-64312-5_44}.

To assess correlations between simulated character traits and learning outcomes, EduAgent~\cite{xu2024eduagent} employed Pearson coefficients, measuring the relationship between demographic/behavioural attributes and performance, thus providing evidence of simulation realism.

For simulations involving exam responses, accuracy was assessed using a custom metric known as M Score~\cite{benedetto2024using}, which evaluated correctness across Simulated Student levels in answering key questions.

In automatic classification tasks, studies employed metrics like F1 Score and Hamming Score to compare models such as SVM and DistilBERT for evaluating student responses, noting that DistilBERT benefited from linguistic feature integration~\cite{schonle2024automatic}.

From an adaptive learning perspective, Agent4Edu~\cite{gao2025agent4edu} used a Computerized Adaptive Testing (CAT) framework, introducing three metrics, satisfaction, appropriateness of difficulty(AoD), and learning gain, to compare personalized learning algorithms from the agent's standpoint.

\begin{table}[h!]
    \centering
\resizebox{\textwidth}{!}{%
    \begin{tabular}{|p{4cm}|p{6cm}|p{6cm}|}
        \hline
        \textbf{Metric Category} & \textbf{Specific Metric(s)} & \textbf{Purpose} \\
        \hline
        \textbf{Performance/Accuracy} & Accuracy, M Score & Evaluating model responses, including correctness across Simulated Student levels \\
        \hline
        \textbf{Utility of AI-Generated Content} & Utility ratings, Manual Annotations & Assessing utility and quality of generated educational content \\
        \hline
        \textbf{Consistency \& Relevance} & Cronbach's Alpha, Student Surveys & Measuring internal consistency and subjective relevance (clarity, depth) \\
        \hline
        \textbf{Programming Feedback} & Precision, Recall, Accuracy, Unit Testing & Validating syntactic and functional correctness of generated feedback \\
        \hline
        \textbf{Semantic \& Information Content} & Semantic Similarity, Information Entropy & Evaluating similarity, richness and informativeness of AI-generated content \\
        \hline
        \textbf{Simulation Realism} & Pearson Coefficients & Measuring correlation between learner traits and performance \\
        \hline
        \textbf{Automatic Classification} & F1 Score & Evaluating model performance on classification tasks \\
        \hline
        \textbf{Adaptive Learning} & AoD, Satisfaction, Suitability of the material, Learning Gain & Assessing user satisfaction, material appropriateness and learning improvement \\
        \hline
    \end{tabular}
}
    \caption{Summary of evaluation metrics for LLMs in education.}
    \label{tab:evaluation_metrics}
\end{table}

\subsection{Survey-Based Evaluations}

Surveys have been used to assess students' experiences and the perceived value of AI-generated content or interactions with LLM-driven systems. These instruments typically explored aspects such as relevance, clarity, depth, and overall satisfaction.

One study used a survey grounded in the Community of Inquiry (CoI) framework~\cite{zhang2024simulating} to evaluate the effect of class roles in simulated classrooms. 

The TeachTune study \citep{jin2024teachtune} engaged ten raters who assessed Simulated Students on a five-point Likert scale, evaluating initial knowledge state, the intensity of four student traits, and authenticity. The raters reported that the Simulated Students behaved as naturally as real students and proved useful for teacher training.

\subsection{Other Evaluation Approaches}

Human evaluation remains a constant in the field, with experts or students assessing the quality, accuracy, and pedagogical relevance of AI-generated content or simulated behaviour. For example, PersonaEval~\cite{zhangpersonaeval} examined the reliability of LLMs in role-playing tasks by categorising responses based on five expertise levels (child, teenager, college student, graduate student, expert).

Some studies compared simulated responses with real student data to assess simulation fidelity~\cite{lu2024generative}, analyzing patterns such as answers to multiple-choice items.

In \textit{Next-step hint generation for introductory programming using large language models}~\cite{roest2024next}, two expert raters qualitatively evaluated hints generated by LLMs based on nine criteria, including feedback type, information, level of detail, personalization, appropriateness, specificity, misleading information, tone, and length.

The article \textit{Investigating the Effect of Role-Play Activity With GenAI Agent on EFL Students' Speaking Performance}~\cite{chen2025investigating} describes a quasi-experimental study involving 53 Chinese students enrolled in an English course. The study included pre- and post-tests to assess speaking performance, intrinsic motivation, and self-efficacy following role-play activities conducted either with a generative AI agent or with classmates.

In \textit{Assessing ChatGPT’s capability for multiple choice questions using RaschOnline: observational study}~\cite{chow2024assessing}, an observational study with 300 simulated participants was analysed using Rasch Modelling to evaluate performance against item difficulty.

Reflective documents voluntarily submitted by students in role-playing experiments also provided qualitative insights into their experiences~\cite{chen2025investigating}.

Finally, studies like MathVC~\cite{yue2024mathvc} conducted ablation studies on prompt design to examine how different strategies influenced behaviour and outputs in a multi-character virtual classroom simulation. Similarly, in studies of hint generation~\cite{tonga2024automatic}, the ability of LLM tutors to detect student errors and self-correct after feedback was analysed.

\section{Discussion and Challenges}

The use of LLMs for student simulation constitutes an emerging field with transformative potential, yet it also presents significant constraints that must be addressed with caution. Three research questions frame the current debate. 

Addressing RQ1, the main cognitive architectures for LLM-based Simulated Students are built upon multi-agent systems and modular components, exemplified by frameworks such as Classroom Simulacra, AICademic, and Agent4Edu. The key implementation mechanism for simulating memory involves a structured management system utilising short- and long-term stores with explicit operations for retrieval, writing, and iterative reflection (e.g., the Transferable Iterative Reflection module in Classroom Simulacra). Reasoning and knowledge Modelling are primarily executed through three strategies: Direct Simulation via Prompts for instantiating learner profiles and cognitive states , Knowledge Tracing for tracking mastery over time , and the use of Knowledge Graphs and Heuristics to provide a structured, interpretable representation of the student's knowledge state.

About RQ2—To what extent do LLM-based systems accurately simulate human behaviour in educational contexts? The literature shows that LLMs generate linguistically coherent interactions and can reproduce basic learning trajectories with acceptable reliability in isolated tasks. Nevertheless, they still fall short of capturing affective variability, spontaneous mistakes and long-term memory management, all of which characterise real learners. Their fidelity therefore sits in an intermediate band: suitable for proof-of-concept studies and low-stakes analyses, but inadequate as a substitute for observations of authentic students in high-resolution or high-risk settings.

A central component of this discussion concerns the practical application and efficacy of the systems reviewed. To facilitate a comparative analysis of the validation methodologies and reported outcomes, Table~\ref{tab:systems_evaluation} synthesises the key systems cited in this work.

The table details the specific role(s) assigned to the Simulated Students, the key performance metrics employed for their assessment, and the nature of any formal educational evaluation conducted. This synthesis provides a foundation from which to discuss current trends, evaluative gaps, and future research directions.

\begin{table}[h!]
\centering
\vspace{1em} 
\resizebox{\textwidth}{!}{

\begin{tabular}{|p{2cm}|p{3.5cm}|p{3cm}|p{3.5cm}|p{4cm}|}
\hline
\textbf{System / Citation} & \textbf{Role(s) of Simulated Student} & \textbf{Key Performance Metrics} & \textbf{Type of Educational Evaluation} & \textbf{Key Outcomes} \\
\hline

Classroom Simulacra \citep{xu2025classroom} & 
Mimics learning behaviours (not cognitive states). & 
Multi-level post-test accuracy & 
Technical validation (no formal educational evaluation). & 
TIR module enhances behavioural simulation accuracy. \\
\hline

Agent4Edu \citep{gao2025agent4edu} & 
Simulator with dynamic cognitive states & 
Agent-based evaluation of learning services; task difficulty. & 
Computerised Adaptive Test (CAT) framework. & 
Compares personalized learning algorithms via agent. \\
\hline

TeachTune \citep{jin2024teachtune} & 
Configurable students (knowledge, traits). & 
Median error: 5\% (knowledge), 10\% (traits). & 
Teacher technical evaluation; user study (30 educators). & 
Aligned with teacher expectations; reduced educator workload. \\
\hline

Generative Students \citep{lu2024generative} & 
Creates learner personas by sampling KCs. & 
MCQ response pattern analysis. & 
Comparison of simulated vs. real student data (fidelity check). & 
High correlation ($r=0.72$) with real data; improved question quality signals. \\
\hline

EduAgent \citep{xu2024eduagent} & 
Generates agents with individual attributes. & 
Pearson coefficients. & 
Correlation: simulated attributes vs. learning outcomes. & 
Provides evidence of realism. \\
\hline

(Benedetto et al. 2024) \citep{benedetto2024using} & 
Simulates MCQ responses by student level. & 
M Score (custom metric). & 
Response correctness evaluated by simulated level. & 
Monotonic accuracy by level; domain generalization (\textbf{not} LLM). \\
\hline

(Chen et al. 2025) \citep{chen2025investigating} & 
GenAI agent for EFL speaking role-play. & 
Pre/post-tests (speaking, motivation, self-efficacy). & 
Quasi-experiment (53 students). & 
GenAI = peers for speaking; GenAI > peers for motivation \& self-efficacy. \\
\hline

MathVC \citep{yue2024mathvc} & 
Multi-character virtual classroom simulation. & 
Surveys (Likert), interviews, log analysis. & 
Empirical study (14 middle-school students). & 
Reported gains in engagement, motivation, confidence. \\
\hline

WIP (Ma and Wang 2024) \citep{ma2024wip} & 
Student Agent gives feedback on materials. & 
Not specified. & 
In-class survey. & 
Survey: improved engagement, comprehension, understanding. \\
\hline

\end{tabular}    
}
\vspace{1em}
\caption{Summary of Simulated Student systems, their roles, performance metrics, and educational evaluation.}
\label{tab:systems_evaluation}
\end{table}

With respect to RQ3—Can these simulations be leveraged effectively to improve teaching practice and instructional design? Evidence from controlled environments (as summarised in Table~\ref{tab:systems_evaluation}) indicates tangible benefits: LLM-generated learners can calibrate assessment items, rehearse tutoring scenarios and provide safe rehearsal spaces for novice teachers without exposing real pupils. Yet their direct pedagogical impact depends on three unresolved conditions alignment with robust educational frameworks, rigorous bias control and broad acceptance by instructors. Until these criteria are satisfied, simulated agents are best employed as analytical adjuncts rather than replacements for human practice.

Lastly, the RQ4—What are the main open research lines in this field? Several interrelated research lines remain open in the simulation of students with LLMs. First, enhancing the cognitive and affective fidelity of simulated agents is essential: current models struggle to reproduce the complexity of real learners’ emotions, misconceptions, and memory dynamics. Second, mitigating algorithmic bias and validating the pedagogical reliability of outputs remain central challenges, especially in high-stakes applications. Third, developing rigorous, multi-modal evaluation frameworks, both human and automatic, is critical to assess the realism, coherence, and educational utility of these agents. Fourth, efforts to simulate individualized learners, incorporating personality traits and socio-emotional dimensions, raise questions about psychological validity and ethical appropriateness. Fifth, the integration of LLMs with symbolic systems and knowledge graphs presents a promising but underexplored direction for neurosymbolic educational agents. Lastly, real-world applications—such as curriculum design, assessment calibration, and teacher training—require scalable deployment and robust empirical validation before widespread adoption can be justified. These areas define the frontier of LLM-based educational simulation and offer a roadmap for future exploration.

As a final synthesis of the main findings, the review demonstrates that the identified simulation roles for LLM agents map directly onto fundamental learning theories, thereby validating their pedagogical and theoretical potential. The role of a 'virtual tutor' or scaffolding agent is primarily grounded in Vygotsky’s ZPD, where the agent acts as a more knowledgeable other (MKO) to provide adaptive support, guidance, and personalized feedback. Similarly, the simulation of 'debate partners' or 'virtual students' supports the collaborative construction of knowledge and the learning-by-teaching approach, by requiring the student-tutor to practice explanations and identify knowledge gaps. Other roles, such as the 'motivator' or the 'expert/stakeholder', facilitate experiential learning by replicating real-world interactions and aligning feedback with motivational theories. In all cases, the efficacy of these simulated roles critically depends on prompt engineering techniques (such as Chain-of-Thought or the detailed definition of personae), which guide the LLM's reasoning process and ensure the agent adheres to the cognitive and behavioural boundaries specific to its simulated role.

\subsection{Ethical Considerations}
The ethical and social dimensions impose a necessary caution on the enthusiasm surrounding these findings. The root of this concern lies in the fact that the corpora underlying LLMs inevitably embed cultural, socioeconomic, and gender biases. The manifestation of these biases in Simulated Students, rather than Modelling real diversity equitably, carries the potential risk of reinforcing pre-existing inequities within the educational system, thereby compromising the impartiality and validity of assessed pedagogical interventions.

Furthermore, the ethical domain is also confronted with serious privacy concerns stemming from data management. The reuse of sensitive educational data without the existence of transparent audit trails—mechanisms necessary to track the origin and utilization of the information—represents a significant risk. This lack of traceability can erode trust and raise fundamental questions regarding confidentiality and data protection upon deploying these tools in real educational contexts.

\subsection{Limitations}
Several limitations temper the enthusiasm surrounding these findings.

Technically, behavioural fidelity remains constrained: models tend to produce idealised answers and struggle with narrative coherence and long-term consistency, which limits their capacity to sustain plausible learning trajectories or to respond reliably to cumulative instruction. This lack of realism jeopardizes the validity of pedagogical interventions evaluated using these Simulated Students, necessitating a focus on mechanisms that model erratic, non-linear, and emotionally rich learning curves.

Methodologically, the field lacks consensus on validation criteria: current practices—dataset comparison, expert judgement and Turing-style tests—are costly, subjective and difficult to scale, while the dominance of Anglophone sources narrows intercultural representativeness and perpetuates epistemic bias. Consequently, the absence of standardized, scalable validation metrics hinders both the cumulative progress of the field and the fair comparison of different LLM-based simulation architectures across global educational contexts.

Ethically and socially, the corpora underlying LLMs embed cultural, socioeconomic and gender biases that can resurface in Simulated Students, potentially reinforcing inequities; moreover, the reuse of sensitive educational data without transparent audit trails raises significant privacy concerns. Therefore, without robust de-biasing strategies and clear ethical guidelines for data provenance, the widespread deployment of these models risks exacerbating digital divides and eroding user trust in AI-driven educational systems.

On the applicative front, integration into real-world platforms (MOOCs, intelligent tutoring systems, virtual-reality classrooms) remains fragmentary: most simulations are confined to laboratory settings, and teacher-training scenarios demand agents whose responses are rich enough in affect and variability to elicit genuine pedagogical reactions, an emotional realism not yet attained by current models. This gap between lab-based proof-of-concepts and real-world utility suggests that the next phase of research must prioritise engineering challenges, focusing on low-latency, ethically aligned, and emotionally expressive agents ready for large-scale, sustained interaction.

\subsection{Open Problems and Research Directions}

Despite recent progress, significant challenges remain. The following list summarizes the outstanding issues and lines of research described throughout this review.

\begin{itemize}
\item \textbf{Grounding internal states.} Methods to observe or verify latent cognitive/affective states (e.g., uncertainty, misconceptions) beyond self-reports or prompted rationales.
\item \textbf{Causal evaluation.} Shift from correlational metrics to causal designs testing whether Simulated Students improve model reasoning, content quality, or human learning outcomes.
\item \textbf{Long-horizon interaction \& memory.} Agents with durable, verifiable memories (weeks/months); study drift, forgetting, and curriculum pacing.
\item \textbf{Pedagogical alignment.} Formalize alignment to instructional principles (mastery learning, formative assessment) and measure adherence, not only task success.
\item \textbf{Controllability and persona stability.} Ensure traits (ability, motivation, misconceptions) are stable and disentangled under prompt/tool changes.
\item \textbf{Safety, privacy, and bias.} Quantify leakage risks with real student data; develop bias audits for demographic/ability groups and fair performance targets.
\item \textbf{Standardized benchmarks.} Public, reproducible suites coupling tasks, rubrics, and human baselines for classroom-like multi-turn interactions.
\item \textbf{Sim-to-real transfer.} Validate that simulated-cohort findings predict outcomes with human learners; define failure modes and correction loops.
\item \textbf{Multi-agent classrooms.} Model peer effects, group work, and teacher orchestration; study coordination, role switching, and conflict resolution.
\item \textbf{Tool-use and multimodality.} Integrate tools (code, search, graders) and modalities (speech, handwriting, diagrams) with traceable attribution of improvements.
\item \textbf{Efficiency and cost.} Report compute/energy per training hour and per simulated learner; design budget-aware algorithms for large-scale cohorts.
\item \textbf{Reproducibility standards.} Release seeds, prompts, agent configs, and eval scripts; adopt reporting checklists tailored to simulated-student studies.
\end{itemize}

\section{Conclusions}

This work has examined the current state of student simulation using LLMs, highlighting both their transformative potential and the main challenges associated with their adoption in educational contexts. The literature review indicates that LLMs are capable of generating plausible educational interactions, assuming student roles with varying knowledge profiles, and participating in simulated classroom dynamics with a considerable degree of realism. These advances open up new avenues for the development of teacher training tools, automated assessment systems, and adaptive learning environments.

Nevertheless, the technical, ethical, and pedagogical challenges must not be underestimated. Significant limitations persist in the behavioural fidelity of simulated agents, the representation of student diversity, and the management of biases inherent to the models. Moreover, empirical validation of the impact of these technologies in authentic educational settings remains limited and calls for more systematic investigation, particularly regarding their formative utility and acceptance by both educators and learners.

The principal contribution of this article is to provide a critical and conceptual foundation to guide future research and technological development in this domain. Specifically, it underscores the need to move towards more robust and transparent simulations, integrated with sound educational frameworks and designed with an inclusive and ethical perspective. Furthermore, it recommends fostering interdisciplinary collaboration and promoting data openness to ensure the responsible advancement of this emerging technology.
 
Ultimately, LLMs represent not only a technical innovation, but also an opportunity to rethink how we design, evaluate, and practise teaching. When appropriately understood and applied, student simulation can become a valuable instrument in the service of more effective, reflective, and human-centred education.

\section*{Competing Interests}
The authors declare that there is no conflict of interest regarding the publication of this paper.

\section*{Acknowledgments}
This work was supported by the Prometeo programme from Conselleria de Innovación, Universidades, Ciencia y Sociedad Digital of Generalitat Valenciana (Spain) under Grant CIPROM/2021/17. Additionally, generative AI tools were utilised as an assistant for drafting and refining the text, ensuring clarity and accuracy in the presentation of this work.

\section*{Statement on the Use of AI}
During the preparation of this work the authors used the LLMs ChatGPT and Google Gemini for two specific purposes. First, they were used to re-evaluate articles excluded during the literature screening process to prevent false negatives. Second, they served to assist in the translation of the full manuscript into academic English. After using this tool/service, the authors reviewed and edited the content as needed and takes full responsibility for the content of the published article.

\bibliographystyle{plainnat}
\bibliography{sn-bibliography}

\end{document}